\documentclass[12pt,preprint,preprintnumbers,a4paper]{article}
\pdfoutput=1

\usepackage{jheppub}

\usepackage{afterpage}

\usepackage{longtable}
\usepackage{multirow}
\usepackage{enumerate}
\usepackage{amsmath}
\usepackage{amsfonts}
\usepackage{amssymb}
\usepackage{graphicx, rotating}
\usepackage{epstopdf}
\usepackage{epsfig}
\usepackage{latexsym}
\usepackage{graphicx}
\usepackage{color}
\usepackage{amsmath,amssymb}
\usepackage{slashed}
\usepackage{hyperref}
\usepackage{wasysym}
\usepackage{MnSymbol}
\usepackage{multirow}
\usepackage{caption}

\definecolor{LightCyan}{rgb}{0.88,1,1}
\definecolor{LightViolet}{rgb}{1, 0.88,1}
\definecolor{LightGreen}{rgb}{0.88, 1, 0.88}
\definecolor{forestgreen}{rgb}{0.13,0.35,0.13}

\definecolor{oucrimsonred}{rgb}{0.6, 0.0, 0.0}
\definecolor{persianblue}{rgb}{0.11, 0.22, 0.73}
\definecolor{forestgreen}{rgb}{0.13,0.35,0.13}
 \hypersetup{colorlinks, citecolor=oucrimsonred, linkcolor=persianblue, urlcolor=oucrimsonred}

\usepackage{color, colortbl}
\definecolor{rossos}{cmyk}{0,1,1,0.55}
\definecolor{bluscuro}{rgb}{0.15, 0.2, .85}
\definecolor{bluchiaro}{cmyk}{1,.3,0.,0.1}
\definecolor{verdechiaro}{rgb}{0.6,.1,0.9}
\definecolor{forestgreen}{rgb}{0.13,0.35,0.13}
\definecolor{gold}{rgb}{1,0.84,0}
\definecolor{darkcyan}{rgb}{0,0.55,0.55}
\definecolor{brown}{rgb}{0.65,0.16,0.16}
\definecolor{orange}{rgb}{1,0.65,0.}

\def\bea{\begin{eqnarray}} \def\eea{\end{eqnarray}}
\def\be{\begin{equation}} \def\ee{\end{equation}}

\newcommand{\promille}{%
  \relax\ifmmode\promillezeichen
        \else\leavevmode\(\mathsurround=0pt\promillezeichen\)\fi}
\newcommand{\promillezeichen}{%
  \kern-.05em%
  \raise.5ex\hbox{\the\scriptfont0 0}%
  \kern-.15em/\kern-.15em%
  \lower.25ex\hbox{\the\scriptfont0 00}}

\begin{document}
\preprint{SISSA/26/2015 FISI}

\vspace*{-30mm}

\title{\boldmath On the stability of the electroweak vacuum\\ in the presence of low-scale seesaw models}

\author[a]{Luigi Delle Rose,}
\author[a]{Carlo Marzo}
\author[b]{and Alfredo Urbano}
\affiliation[a]{Dipartimento di Matematica e Fisica ``Ennio De Giorgi", Universit\`a del Salento and INFN, sez. di Lecce. Via Arnesano, 73100 Lecce, ITALY.}
\affiliation[b]{SISSA - International School for Advanced Studies, via Bonomea 265, I-34136, Trieste, ITALY.}

\emailAdd{luigi.dellerose@le.infn.it}
\emailAdd{carlo.marzo@le.infn.it}
\emailAdd{alfredo.urbano@sissa.it}

\vspace{2cm}


\abstract{
The scale of neutrino masses and the Planck scale are separated by more than twenty-seven order of magnitudes. 
However, they can be linked by imposing the stability of the electroweak (EW) vacuum.
The crucial ingredient is provided by the generation of neutrino masses via a seesaw mechanism
triggered by Yukawa interactions between the standard model (SM) Higgs and lepton doublets and additional heavy right-handed neutrinos.
These neutrinos participate to the renormalization group (RG) running of the dimensionless SM
couplings, affecting their high-energy behavior. 
The Higgs quartic coupling is dragged towards negative values, thus altering the stability  of the EW vacuum.
In the usual type-I seesaw model, this effect is too small to be a threat since, in order to comply with low-energy neutrino data, one is forced to consider 
either too small Yukawa couplings or too heavy right-handed neutrinos.
In this paper we explore this general idea in the context of low-scale seesaw models.
These models are characterized by sizable Yukawa couplings and right-handed neutrinos with mass of the order of the EW scale, thus maximizing their
impact on the RG flow. 
As a general result, we find that Yukawa couplings such that ${\rm Tr}(Y_{\nu}^{\dag}Y_{\nu}) \gtrsim 0.4$ are excluded. 
We discuss the impact of this bound on several observables, with a special focus on 
the lepton flavor violating process $\mu\to e\gamma$ and the neutrino-less double beta decay.
}
\maketitle

\section{Introduction}\label{sec:Intro}
The discovery of the Higgs boson at the LHC~\cite{Aad:2012tfa,Chatrchyan:2012ufa} represents 
a cornerstone in particle physics, finally achieved after a nearly half-century quest.
Paradoxically though it may seem, this astonishing discovery did not clarify the true origin of the electroweak (EW) symmetry breaking: after LHC run I, 
any clue of new physics beyond the Standard Model (SM) is still missing.

To crown it all, the measured value of the Higgs boson mass, $M_h = 125.09\pm 0.21_{{\rm stat.}} \pm 0.11_{{\rm syst.}}$ GeV 
according to the latest combination of both ATLAS and CMS experiments~\cite{Aad:2015zhl}, 
is consistent, in spite of any naturalness argument, even if the SM is extrapolated up to the Planck scale. 
This remarkable conclusion is based 
on the fact that  the value of the Higgs boson mass 
is caught in the vice of two solid theoretical constraints if one tries to extend the SM up to arbitrarily high energies by means 
of its Renormalization Group Equations (RGEs). On the one hand, for large values of $M_h$, 
the Higgs quartic coupling $\lambda$ is driven towards greater and greater values, eventually becoming non-perturbative. 
On the other one, for small values of $M_h$, $\lambda$ is driven towards smaller values, eventually becoming negative.
The occurrence of the latter condition may have dramatic  consequences
since it implies that  
the EW vacuum is only a local minimum, and there exists a global minimum at higher energy in the Higgs potential~\cite{Krive:1976sg,Krasnikov:1978pu,Maiani:1977cg,Politzer:1978ic,Hung:1979dn,Cabibbo:1979ay,Linde:1979ny,Lindner:1985uk,Lindner:1988ww,Sher:1988mj,Arnold:1989cb,Arnold:1991cv,Sher:1993mf,Altarelli:1994rb}.
For a sufficiently old Universe, nothing would prevent the vacuum expectation value (vev) of the Higgs field
to be at the global minimum, as a mere consequence of a tunneling transition between the two vacua~\cite{Coleman:1977py,Callan:1977pt,Coleman:1980aw}.
Fortunately, Nature spared us from this unspeakable catastrophe: in the SM with $M_h \simeq 125$ GeV
 $\lambda$ does become negative at high energies, but its absolute value remains far too small to trigger the tunneling. 
 In other words, in the SM the EW vacuum is metastable~\cite{Holthausen:2011aa,EliasMiro:2011aa,Alekhin:2012py}. 
 
The metastability of the EW vacuum  holds under the assumption that  the SM is the full theory all the way up to the Planck scale.
This argument poses a legitimate question: what is the role of new physics beyond the SM with respect to the stability of the EW vacuum~\cite{EliasMiro:2012ay,Lebedev:2012zw,Bezrukov:2012sa,Datta:2012db,Anchordoqui:2012fq,Masina:2012tz,Kobakhidze:2013pya,Kobakhidze:2014xda,Kobakhidze:2013tn,Chao:2012mx,Chao:2012xt,Moha}?
 First and foremost, to address this question we have to carefully specify which kind of new physics we are referring to.
Undeniably, the most compelling case is provided by the observation of non-zero neutrino masses and oscillations. 
This fact cries out for the inclusion in the SM of a mechanism responsible for the generation of neutrino masses~\cite{Weinberg:1980bf}.
A very economical way is provided by the so-called type-I seesaw mechanism~\cite{seesaw1,seesaw2,seesaw3,seesaw4}.
In a nutshell, the SM is enlarged with the addition of three heavy
 right-handed neutrinos\footnote{An even simpler setup would imply only two sterile neutrinos~\cite{Frampton:2002qc}.} with a Majorana mass $M_R$ breaking lepton number, 
 coupled to the SM Higgs doublet $H$ and leptonic doublets $L$ via a Yukawa coupling $Y_{\nu}$.
 Integrating out the heavy fields, the effective light neutrino mass $m_{\nu} \approx v^2 Y_{\nu}/M_R$ is generated after EW symmetry breaking,  
 where $v \simeq 246$ GeV is the EW vev.
 
 In principle, this additional Yukawa interaction pushes the SM towards the instability region, since, at least at one-loop, 
 it exactly mimics the  effect of the top Yukawa coupling on the running of $\lambda$: 
 it introduces a negative contribution which drags the Higgs quartic coupling towards negative values~\cite{Casas:1999cd,EliasMiro:2011aa,Chen:2012faa,Rodejohann:2012px}. 
 However, in order to have a sizable effect,  the Yukawa coupling $Y_{\nu}$ must be of order one. 
This is a necessary but not a sufficient requirement.
Indeed, it implies  $M_R \sim \mathcal{O}(10^{15})$ GeV if we assume $m_{\nu} \simeq 0.1$ eV for the light neutrino mass scale.
 This means that the right-handed neutrinos actively participate to the running of $\lambda$ only for values of the RG scale larger than $M_R$.
 Or, to put it another way, there is not enough time, in terms of RG evolution, 
 to sizably alter the SM picture. 
 This result does not change trying to lower the mass scale $M_R$ since in this case,
  in order to reproduce the correct order of magnitude for the mass scale of light neutrinos, one is forced to consider 
$Y_{\nu} \sim \mathcal{O}(10^{-5})$. Of course, this value is too small to modify the running of $\lambda$.

However, this is anything but the end of the story. 
Despite its simplicity, the minimal type-I seesaw suffers from a penalizing phenomenological issue:
 it is not testable. As discussed before, it introduces either unaccessible large mass scale or extremely tiny Yukawa couplings.
The necessity to overcome this unpleasant problem led to the introduction of extended seesaw models with both TeV-scale  right-handed neutrinos 
and sizable Yukawa couplings~\cite{Mohapatra:1986bd,GonzalezGarcia:1988rw,Pilaftsis:1991ug,Mohapatra:2005wg,deGouvea:2006gz,Kersten:2007vk}.
The key feature of these models is that lepton number is softly broken via the introduction of extra singlet fermions in addition to the usual right-handed neutrinos. 
The mass $M_R$ can be brought down to the EW scale without neither causing problem with low-energy neutrino phenomenology nor lowering the Yukawa coupling $Y_{\nu}$.
As a byproduct of this construction, a very rich low-energy phenomenology emerges. 
Potentially interesting signals include, for instance, lepton flavor violating radiative decays, deviations from EW precision observables, and production at colliders of heavy Majorana fermions. 

As already clear from these introductory comments, 
low-scale seesaw models may alter the metastability of the EW vacuum since they feature, at the same time, sizable Yukawa couplings and relatively low mass thresholds~\cite{Khan:2012zw}.
In this work we study in detail the stability of the EW vacuum considering the so-called inverse seesaw model (ISS)~\cite{Mohapatra:1986bd,GonzalezGarcia:1988rw} as a reference model with low-scale seesaw. We include in our analysis the constraints coming from low-energy neutrino phenomenology
with the aim to provide a complete and realistic description of the physics involved.

Schematically, this paper is organized as follows.
In section~\ref{sec:Neutrino} we discuss the most relevant features of the ISS model, with a special focus on the phenomenological constraints 
included in our analysis. 
In section~\ref{sec:Stability} we study the stability of the EW vacuum in the ISS model, 
using the results of section~\ref{sec:Neutrino}.
In section~\ref{sec:Results} we present our results, and finally we conclude 
in section~\ref{sec:Conclusions}. In appendix~\ref{sec:AppA} we generalize our results to the linear and double seesaw models.
 
\section{The inverse seesaw model}\label{sec:Neutrino}

In this section we start our analysis describing the most relevant prerogatives of the ISS model. 
In section~\ref{sec:Model} we introduce the Lagrangian, the mass matrix and the corresponding diagonalization. In section~\ref{sec:BoundsLowEnergy} we discuss 
the constraints coming from low-energy neutrino phenomenology included in our analysis while in section~\ref{sec:NumericalSetup} we
present our strategy for the numerical analysis. 

\subsection{The model}\label{sec:Model}

In the ISS model the SM field content is extended to incorporate 
$n_R$ right-handed neutrinos $N_R^i$ and 
$n_S$  singlet fermionic fields $S^i$.
The Lagrangian in the ISS model is given by
\begin{equation}
\label{eq:LISS}
\mathcal{L}_{\rm ISS} = 
i\overline{N_R}\gamma^{\mu}(\partial_{\mu}N_R) + i\overline{S}\gamma^{\mu}(\partial_{\mu}S)
-\left[\overline{N_R}Y_{\nu}\tilde{H}^{\dag}L + \overline{N_R}M_R S + \frac{1}{2}\overline{S^C}\mu_S S +  h.c.\right]~,
\end{equation}
where $L \equiv (L^e, L^\mu, L^\tau)^T$
represents the left-handed lepton doublets with the usual contents $L^{l=e,\mu,\tau} = (\nu_{l L}, l_L)^T$ while $H$ is the Higgs field, with $\tilde{H}\equiv i\sigma_2 H^*$. 
$Y_{\nu}$ is the  $n_R\times 3$  Yukawa matrix mediating the interactions between the SM leptons and the right-handed neutrinos while $M_R$ and $\mu_S$ are, respectively, $n_R \times n_S$ and $n_S \times n_S$ mass matrices.
Both right-handed neutrinos and singlet fermions have lepton number $\mathbb{L} = 1$; consequently, the mass term  $\overline{S^C}\mu_S S$ violates lepton number for two units.

Introducing the left-handed basis $N_L \equiv (\nu_L, N_R^{\,C}, S)^T$ we have, after EW symmetry breaking, the following mass matrix
\begin{equation}\label{eq:MasterMassMatrix}
\mathcal{M} =
\left(
\begin{array}{ccc}
 0  & m_D^T  & 0  \\
 m_D &  0 & M_R  \\
 0 & M_R^T  &  \mu_S
\end{array}
\right)~,
\end{equation}
with $m_D \equiv vY_{\nu}/\sqrt{2}$.

The mass matrix in eq.~(\ref{eq:MasterMassMatrix}) can be diagonalized by means of the following unitary transformation
\begin{equation}\label{eq:UnitaryU}
U^{T}\mathcal{M} U = \mathcal{M}_{\Delta} ~~\Rightarrow~~
U^{\dag}(\mathcal{M} ^{\dag}\mathcal{M})U = \mathcal{M}_{\Delta}^2~,~~~~N^{\prime}_L = U^{\dag}N_L~,
\end{equation}
where $\mathcal{M}_{\Delta}$ is the diagonal matrix referred to the mass eigenstates $N^{\prime}_L$. 
The first three eigenstates correspond to the standard light active neutrinos while the remaining $n_R + n_S$ states are additional heavy sterile neutrinos.

Following the standard seesaw calculation~\cite{Schechter:1981cv,Forero:2011pc} and assuming the hierarchy $M_R\gg m_D \gg \mu_S$, it is possible to extract 
the effective light neutrino mass matrix
\begin{equation}\label{eq:LightNeutrinoMass}
m_{\nu} \approx m_D^T\left(M_R^{\,T}\right)^{-1}\mu_S M_R^{\,-1}m_D~.
\end{equation}

In the following we consider the case $n_R = n_S = 3$.  Moreover, without loss of generality, we take $M_R$ to be real and diagonal.
We also work in a basis in which the mass matrix of charged SM leptons is diagonal. 
Within this framework, in the $\mu_S \to 0$ limit the three light neutrinos are massless while the six heavy neutrinos can be recast into three pairs of Majorana particles with three (double degenerate) masses $M_{Ri}$, $i=1,2,3$.
On a general ground, from eq.~(\ref{eq:LightNeutrinoMass}) it follows that the order of magnitude 
of the light neutrino mass is $m_{\nu} \sim \mathcal{O}(\mu_S \times m_D^2/M_R^2)$. Assuming $\mu_S \sim \mathcal{O}(1)$ keV, the model can accommodate 
sub-eV light neutrino masses with $Y_{\nu} \sim \mathcal{O}(1)$  couplings and $M_R\sim \mathcal{O}(1-10)$ TeV seesaw scale.
These order of magnitude estimates lie at the hearth of the ISS scenario. Small values of $\mu_S$ are expected by virtue of the 't Hooft naturalness criterium~\cite{Naturalezza}, 
since the limit $\mu_S\to 0$ increases the symmetry of the theory. 
Interestingly, the keV scale nicely fits the typical mass scale characterizing warm dark matter; in the ISS models with $n_R \neq n_S$ the spectrum -- in addition to light and heavy neutrinos -- also contains intermediate states with keV mass that are valuable warm dark matter candidates~\cite{Abada:2014zra}.
The estimates $Y_{\nu} \sim \mathcal{O}(1)$ and $M_R\sim \mathcal{O}(1-10)$ TeV represent the most relevant phenomenological properties of the ISS model 
since they allow for sizable (and, in principle, measurable) mixing effects between light active and heavy sterile neutrinos.
This issue is particularly striking if compared with the typical high-scale characterizing the minimal type-I seesaw~\cite{seesaw1,seesaw2,seesaw3,seesaw4} -- that is $M_R \sim \mathcal{O}(10^{15})$ GeV for order one Yukawa couplings -- in which 
mixing effects, typically of order $\mathcal{O}(m_D^2/M_R^2)$, are negligible.
As we shall explain in detail in section~\ref{sec:Stability}, the occurrence 
of both the peculiar conditions $Y_{\nu} \sim \mathcal{O}(1)$ and $M_R\sim \mathcal{O}(1-10)$ TeV is of fundamental importance to determine 
the stability of the EW vacuum in the context of the ISS model.

Of particular relevance for many phenomenological applications is the generalized Casas-Ibarra parametrization~\cite{Casas:2001sr}
\begin{equation}\label{eq:CasasIbarra}
Y_{\nu} = \frac{\sqrt{2}}{v} V^* \sqrt{\hat{M}}  R  \sqrt{\hat{m}_{\nu}} U^{\dag}_{\rm PMNS}~,
\end{equation}
where $\sqrt{\hat{m}_{\nu}}$ is the diagonal matrix defined by the square roots of the eigenvalues corresponding to the three light neutrinos, $m_{\nu i}$ with $i=1,2,3$ hereafter, and 
$\sqrt{\hat{M}}$ is the diagonal matrix containing the square roots of the eigenvalues of $M \equiv M_R \mu_S^{\,-1}M_R^{\,T}$ whose 
diagonalization is defined by means of the transformation $V^TMV = \hat{M}$. 
$R$ is an arbitrary $3\times 3$ complex orthogonal matrix parametrized by three complex angles which encodes the remaining degrees of freedom.
Finally, $U_{\rm PMNS}$ corresponds to the unitary 
Pontecorvo-Maki-Nakagawa-Sakata (PMNS) leptonic  mixing matrix. Assuming the standard picture with three neutrino flavors the matrix $U_{\rm PMNS}$ can be parametrized as follows
\begin{equation}\label{eq:PMNS}
U_{\rm PMNS} =
\left(
\begin{array}{ccc}
 c_{12}c_{13} & s_{12}c_{13}  & s_{13}e^{-i\delta_{\rm CP}}  \\
 -s_{12}c_{23}-c_{12}s_{13}s_{23}e^{i\delta_{\rm CP}} & c_{12}c_{23}-s_{12}s_{13}s_{23}e^{i\delta_{\rm CP}}  & c_{13}s_{23}  \\
 s_{12}s_{23} - c_{12}s_{13}c_{23}e^{i\delta_{\rm CP}} & -s_{12}s_{23} - s_{12}s_{13}c_{23}e^{i\delta_{\rm CP}}  & c_{13}c_{23}  
\end{array}
\right)~,
\end{equation}
where $c_{ij}\equiv \cos\theta_{ij}$ and $s_{ij}\equiv \sin\theta_{ij}$. 
In addition to the Dirac CP violation phase $\delta_{\rm CP}$ there are also two Majorana CP violation phases (not shown in eq.~(\ref{eq:PMNS})). 
The latter are physical only if light neutrinos are Majorana particles, otherwise they can be always rotated away from the Lagrangian in the mass basis. 
In the following we omit the three phases of the PMNS matrix, 
since the Majorana phases are completely unknown and there are only preliminary hints about a non-zero Dirac phase.
 
\subsection{Bounds from low-energy neutrino data}\label{sec:BoundsLowEnergy}

Global analysis of neutrino oscillation data, based on the latest results of the Daya Bay~\cite{An:2013zwz}, RENO~\cite{Ahn:2012nd,RENOtalk}, T2K~\cite{Abe:2013fuq,Abe:2013hdq} and MINOS~\cite{Adamson:2013whj,Adamson:2013ue} experiments, 
allowed to determine the oscillation parameters $\Delta m_{21}^2$, $|\Delta m_{31}^2|$ ($|\Delta m_{32}^2|$, depending on the ordering), $\theta_{12}$, $\theta_{23}$, $\theta_{13}$  with unprecedented high precision, thus opening the era of neutrino precision measurements. 

In this work, we use the latest results of the {\Large $\nu$}fit group~\cite{Gonzalez-Garcia:2014bfa}. As customary, we define $\Delta m_{ij}^2\equiv m_{\nu i}^2 - m_{\nu j}^2$, 
and we adopt the convention that results for the mass squared differences are reported with respect to the one with the largest absolute value.

\begin{itemize}

\item[$\circ$] \textbf{Neutrino mass squared differences}

3-${\sigma}$ C.L. ranges on the mass squared differences 

\begin{equation}
\Delta m_{21}^2/10^{-5}{\rm eV}^2  = (7.02 \to 8.09)~,~~~
\left\{
\begin{array}{c}
  \Delta m_{31}^2/10^{-3}{\rm eV}^2  = (2.317 \to 2.607)~~~~~~{\rm NO}   \\
 \Delta m_{32}^2/10^{-3}{\rm eV}^2  = (-2.590 \to -2.307)~~~{\rm IO}
\end{array}
\right.
\end{equation}
where the first (second) possibility refers to the assumption of normal (inverted) ordering.

\item[$\circ$] \textbf{Leptonic mixing matrix}

3-${\sigma}$ C.L. ranges on the magnitude of the elements of the leptonic mixing matrix in eq.~(\ref{eq:PMNS})
\begin{equation}
\sin^2\theta_{12} = (0.270 \to 0.344)~,
\end{equation}
\begin{equation}
\sin^2\theta_{23} = \left\{
\begin{array}{c}
(0.382 \to  0.643) \\
(0.389      \to  0.644)
\end{array}
\right.~,~~~
\sin^2\theta_{13} = \left\{
\begin{array}{c}
 (0.0186 \to  0.0250)~~~~~{\rm NO} \\
 (0.0188 \to 0.0251)~~~~~{\rm IO}
\end{array}
\right.
\end{equation}

\item[$\circ$] \textbf{Unitarity}

In  the  basis  where  the  charged  lepton  mass  matrix  is  diagonal,  the  leptonic  mixing
matrix in the ISS model is given by the rectangular $3\times 9$ sub-matrix corresponding to the first three rows of the matrix  $U$ defined in eq.~(\ref{eq:UnitaryU}), with the $3\times 3$ block corresponding to the (non-unitary) $\tilde{U}_{\rm PMNS}$. 
Bounds on the non-unitarity of the matrix $\tilde{U}_{\rm PMNS}$ were derived in~\cite{Antusch:2008tz,Antusch:2014woa,Basso:2013jka} using an effective field theory approach.
These bounds can be recast as follows\footnote{Strictly speaking, the bounds in eq.~(\ref{eq:UnitarityBounds}) are valid only if the masses of the sterile neutrinos lie above the EW scale (where they can be safely integrated out). This is always the case in our numerical analysis (see section~\ref{sec:NumericalSetup}).}
\begin{eqnarray}\label{eq:UnitarityBounds}
\epsilon_{\alpha\beta} \equiv \left| 
\sum_{i=4}^{9}U_{\alpha i}U^*_{\beta i}
\right|
&=& \left|
\delta_{\alpha\beta} - (\tilde{U}_{\rm PMNS}\tilde{U}_{\rm PMNS}^{\dag})_{\alpha\beta}
\right|~, \\
\left|
\tilde{U}_{\rm PMNS}\tilde{U}_{\rm PMNS}^{\dag}
\right| &=&
\left(
\begin{array}{ccc}
(0.9979 \to 0.9998)  &  <\,10^{-5} & <\,0.0021  \\
<\,10^{-5}  & (0.9996\to 1.0) & <\,0.0008  \\
 <\,0.0021 & <\,0.0008  &  (0.9947\to 1.0)  
\end{array}
\right)~.\nonumber
\end{eqnarray}

\item[$\circ$]  \textbf{Additional constraints}

The absolute values of neutrino masses $m_{\nu i}$ are unknown. Cosmology sets the most stringent upper bounds using data from the Cosmic Microwave Background (CMB) radiation, supernovae and galaxy clustering. Assuming the validity of the $\Lambda$CDM model~\cite{Ade:2013zuv}, the Planck collaboration placed the upper bound 
$\sum_{i} m_{\nu i} < 0.66$ eV at 95\% C.L.~\cite{Ade:2013zuv}; this bound becomes even more stringent adding data on the Baryon Acoustic Oscillation, $\sum_{i} m_{\nu i} < 0.23$ eV at 95\% C.L.~\cite{Ade:2013zuv}. In our analysis we scan over the interval $10^{-4}\,{\rm eV} \leqslant m_{\nu 1} \leqslant 10^{-1}\,{\rm eV}$ for the mass of the lightest neutrino.

\end{itemize}

\subsection{Relevant parameter space and setup for the numerical analysis}\label{sec:NumericalSetup}

\subsubsection{Target observables}\label{sec:TargetObservables}

The presence of sterile neutrino states affects the SM charge current interaction via the mixing matrix in eq.~(\ref{eq:UnitaryU}). Going from gauge to mass eigenstates
we have
\begin{equation}
\mathcal{L}_{\rm CC} = -\frac{g_2}{\sqrt{2}}\sum_{l = e,\mu,\tau}\overline{l_L}\gamma^{\mu}W^-_{\mu}\nu_{l L} + h.c. = 
-\frac{g_2}{\sqrt{2}}\sum_{l = e,\mu,\tau}\sum_{i = 1}^9
\overline{l_L}\gamma^{\mu}W^-_{\mu}
U_{li}N_L^{\prime\, i}  + h.c.~,
\end{equation}
where $g_2$ is the weak coupling constant.
By means of these interactions, 
and depending on their masses and mixings with light active neutrinos, the presence of new sterile states 
can sizably affect numerous observables, like for instance leptonic and semi-leptonic decays (with a special focus on flavor violating processes)~\cite{Shrock:1980vy,Shrock:1980ct,Ilakovac:1994kj,Deppisch:2004fa,Abada:2012mc,Abada:2013aba,Dinh:2012bp,Ibarra:2011xn,Ibarra:2010xw,Humbert:2015epa}, invisible $Z$ boson decay width~\cite{Akhmedov:2013hec}, Higgs boson decays~\cite{BhupalDev:2012zg,Das:2012ze,Cely:2012bz,Bandyopadhyay:2012px}, 
direct production in meson decay~\cite{Kusenko:2009up}.

In this paper, in order to investigate 
the stability of the EW vacuum
in a region of the parameter space of particular interest for present and future experimental prospects, we focus on the 
lepton flavor violating process $\mu \to e\gamma$ and the neutrino-less double beta decay ($0\nu 2\beta$ hereafter). 

As far as the radiative $\mu \to e\gamma$ decay is concerned, the rate induced by the presence of sterile neutrinos is given by~\cite{Petcov:1976ff,Bilenky:1977du}
\begin{equation}\label{eq:BrLFV}
{\rm Br}(\mu \to e\gamma) = \frac{3\alpha}{32\pi}\left|
\sum_{i=1}^9 U_{\mu i}^*U_{ei}\,\mathcal{G}\left(\frac{m_{\nu i}^2}{M_W^2}\right)
\right|^2~,
\end{equation}
where $\alpha$ is the electromagnetic fine structure constant and $M_W$ the $W$ mass. The loop function is given by $\mathcal{G}(x) \equiv (10 - 43 x +78 x^2 -49 x^3
+ 4 x^4 +18 x^3 \ln x)/[3(-1+ x)^4]$. The present experimental upper bound, reported by the MEG collaboration, is ${\rm Br}(\mu \to e\gamma) < 5.7 \times 10^{-13}$ at 90\% C.L.~\cite{Adam:2013mnn}. 

The amplitude of the  $0\nu 2\beta$ process is proportional to the so-called effective neutrino mass, $m_{\rm eff}^{\nu_e}$. 
Current experiments (among others, GERDA~\cite{Agostini:2013mzu}, EXO-200~\cite{Auger:2012ar,Albert:2014awa}, and KamLAND-ZEN~\cite{Gando:2012zm}), put an upper limit in the range $|m_{\rm eff}^{\nu_e}| \lesssim 140 - 700$ MeV. In the presence of sterile state the effective neutrino mass is given by~\cite{Blennow:2010th}
\begin{eqnarray}\label{eq:EffectiveNeutrinoMass}
m_{\rm eff}^{\nu_e} &\simeq& \sum_{i = 1}^9 U_{ei}^2\,p^2 \frac{m_{\nu i}}{p^2 - m_{\nu i}^2} 
\simeq \sum_{i=1}^3 U_{ei}^2\,m_{\nu i} + p^2\left(
U_{e4}^2\frac{m_{\nu 4}}{p^2 - m_{\nu 4}^2}
+U_{e5}^2\frac{m_{\nu 5}}{p^2 - m_{\nu 5}^2}
\right. 
\nonumber \\
&+&
 \left.U_{e6}^2\frac{m_{\nu 6}}{p^2 - m_{\nu 6}^2}  +U_{e7}^2\frac{m_{\nu 7}}{p^2 - m_{\nu 7}^2}
+U_{e8}^2\frac{m_{\nu 8}}{p^2 - m_{\nu 8}^2} + U_{e9}^2\frac{m_{\nu 9}}{p^2 - m_{\nu 9}^2}\right)~,
\end{eqnarray}
where $p^2 \simeq -(125\,{\rm MeV})^2$ is the momentum of the virtual neutrino.
Notice that, since we are considering the regime $m^2_{\nu i=4,\dots, 9} \gg p^2$, heavy neutrinos decouple in eq.~(\ref{eq:EffectiveNeutrinoMass}) and
the  dominant contribution to $m_{\rm eff}^{\nu_e}$ comes from the light active neutrinos.

\subsubsection{Strategy and first numerical results}\label{sec:NumericalSetup2}

\begin{figure}[!htb!]
\minipage{0.5\textwidth}
  \includegraphics[width=1.\linewidth]{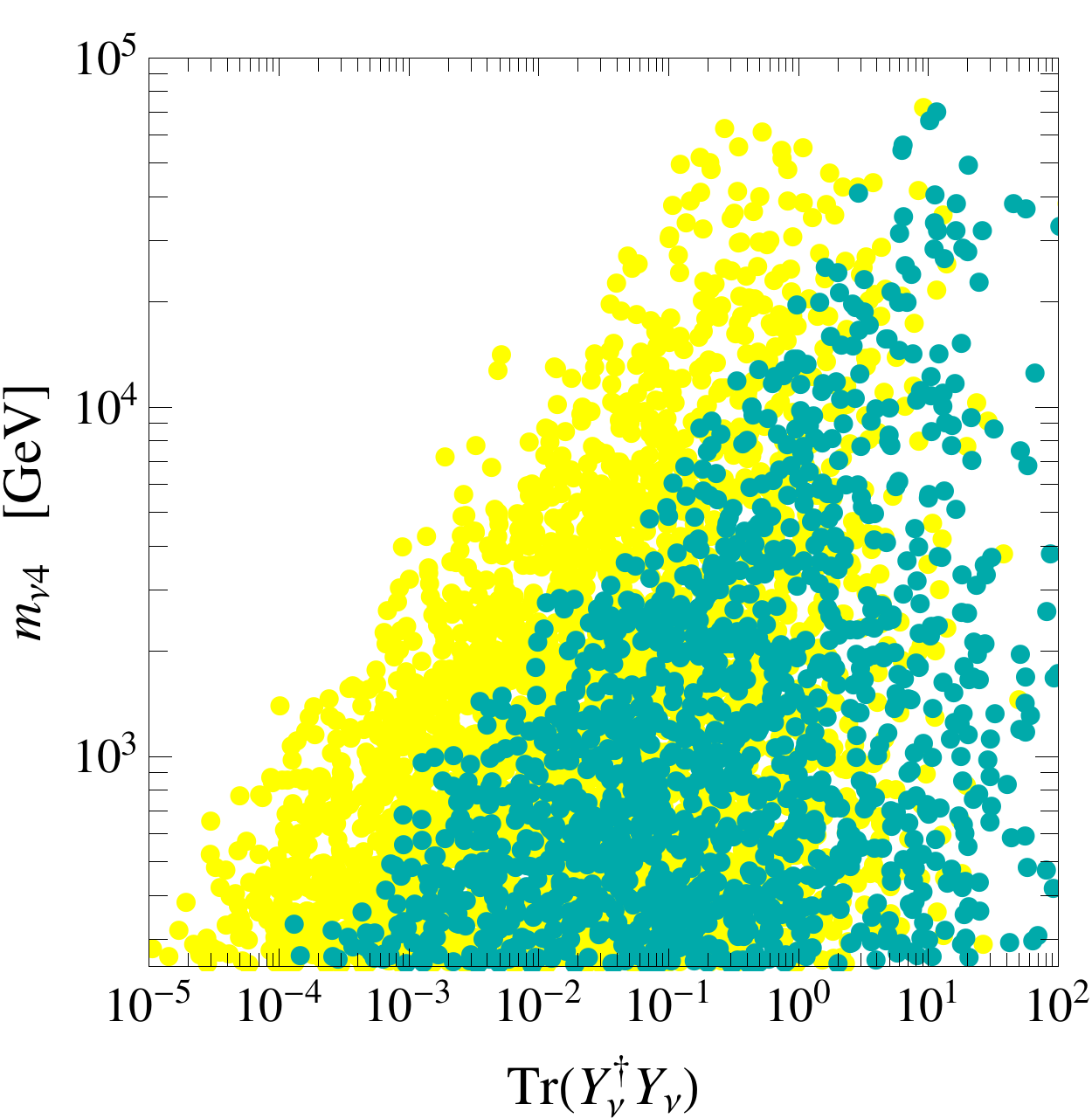}
\endminipage\hfill
\minipage{0.5\textwidth}
  \includegraphics[width=1.\linewidth]{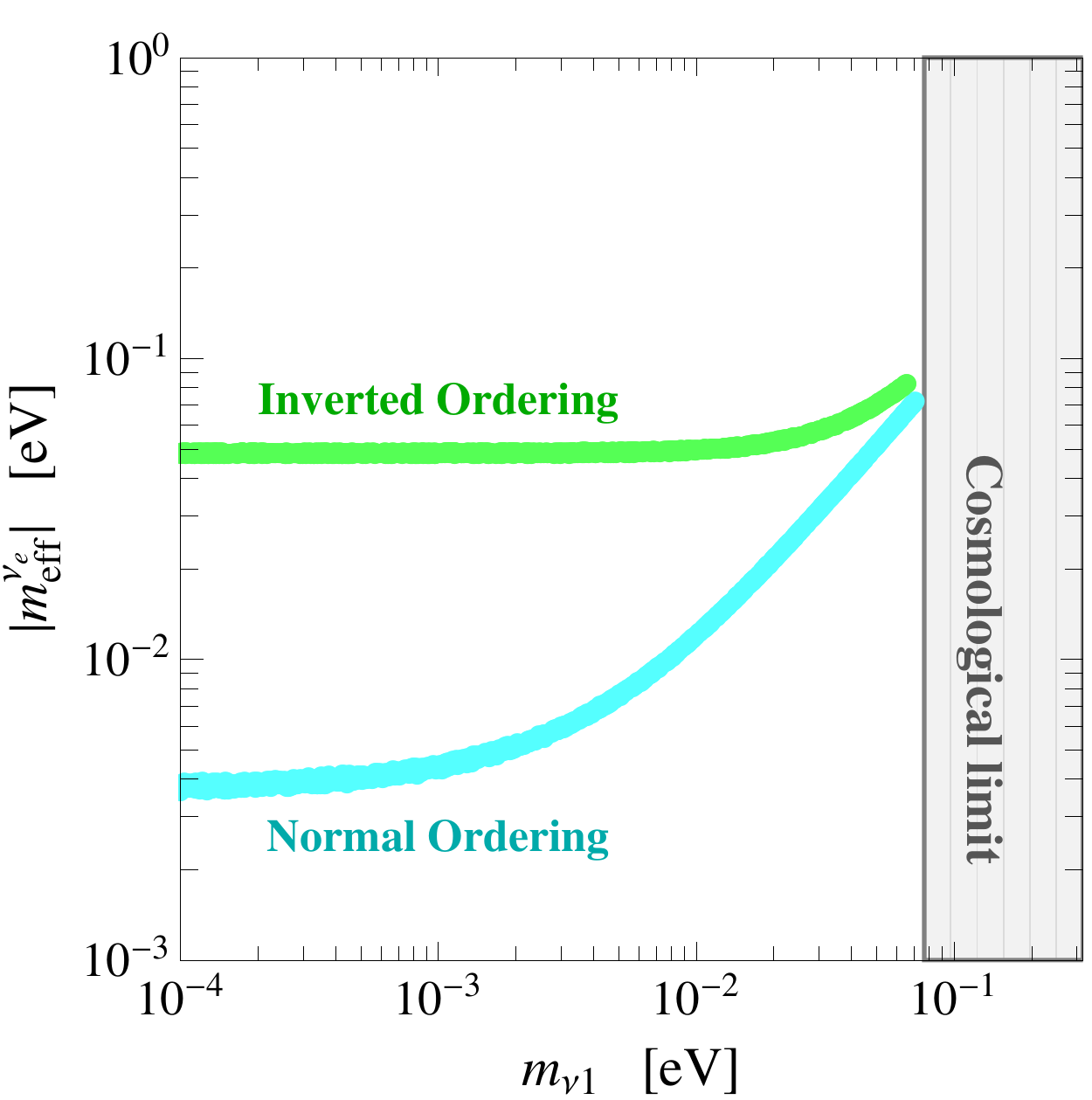}
\endminipage\\
\caption{\it Left panel. 
Distribution of our numerical scan in the plane $[{\rm Tr}(Y^{\dag}_{\nu}Y_{\nu}), m_{\nu 4}]$ considering 
the normal ordering (the inverted ordering gives an analogous result). All points comply with the bounds discussed in section~\ref{sec:BoundsLowEnergy}.
Points with 
${\rm Br}(\mu \to e\gamma) \geqslant 10^{-13}$ are marked in dark cyan. 
On the qualitative level, 
the plot shows that Yukawas ${\rm Tr}(Y^{\dag}_{\nu}Y_{\nu})\gtrsim 0.5$ arise in the whole range of analyzed masses for the right-handed neutrinos. Right panel. Effective neutrino mass as a function of the lightest neutrino mass considering both normal and inverted ordering. 
 Our scan correctly reproduces, as expected if the heavy neutrinos decouple if compared with the typical virtual momentum $p^2 \simeq -(125\,{\rm MeV})^2$, the well-known result characterizing the presence of only three light active neutrinos.
}\label{fig:Temp}
\end{figure}

We perform a scan over the parameter space of the model, and our procedure goes as follows.
First, we randomly generate  {\it i)}  the light neutrino masses $m_{\nu i=1,2,3}$ and the leptonic mixing angles $\theta_{12}$, $\theta_{23}$, $\theta_{13}$ according to
 the corresponding 1-$\sigma$
intervals allowed by the analysis of present experimental data,  {\it ii)} the entries of the matrices $M_{R}$ and $\mu_S$ in the intervals $10^2\,{\rm GeV}\leqslant M_{R i} \leqslant 10^2\,{\rm TeV}$,
 $10^{-1}\,{\rm keV}\leqslant (\mu_S)_{ij} \leqslant 10^2\,{\rm keV}$ and  {\it iii)}  the complex angles defining the arbitrary matrix $R$ in the interval $[0, 2\pi]$.
 Second, we reconstruct the full Yukawa matrix $Y_{\nu}$ using the generalized Casas-Ibarra parametrization in eq.~(\ref{eq:CasasIbarra}).
 Finally, plugging back $M_R$, $\mu_S$ and $Y_{\nu}$ into eq.~(\ref{eq:MasterMassMatrix}), we diagonalize the mass matrix $\mathcal{M}$ in 
 order to find the full $9\times 9$ mixing matrix $U$. 
 The phases of the mixing matrix are fixed using eq.~(\ref{eq:UnitaryU}), by means of the condition $m_{\nu i}\geqslant 0$ for all $i$.
 As a consistency test, for each point of the scan we check  that the mass matrix $\mathcal{M}$, randomly generated as discussed above, 
 correctly reproduces after diagonalization light neutrino masses and mixing angles in agreement with the bounds discussed in section~\ref{sec:BoundsLowEnergy}.
Equipped by these results we can easily compute the 
branching ratio ${\rm Br}(\mu \to e\gamma)$ in eq.~(\ref{eq:BrLFV}) and the effective neutrino mass in eq.~(\ref{eq:EffectiveNeutrinoMass}).
\begin{figure}[!htb!]
\begin{center}
\fbox{\footnotesize NORMAL ORDERING}
\end{center}
\vspace{-0.5cm}
\minipage{0.5\textwidth}
  \includegraphics[width=1.\linewidth]{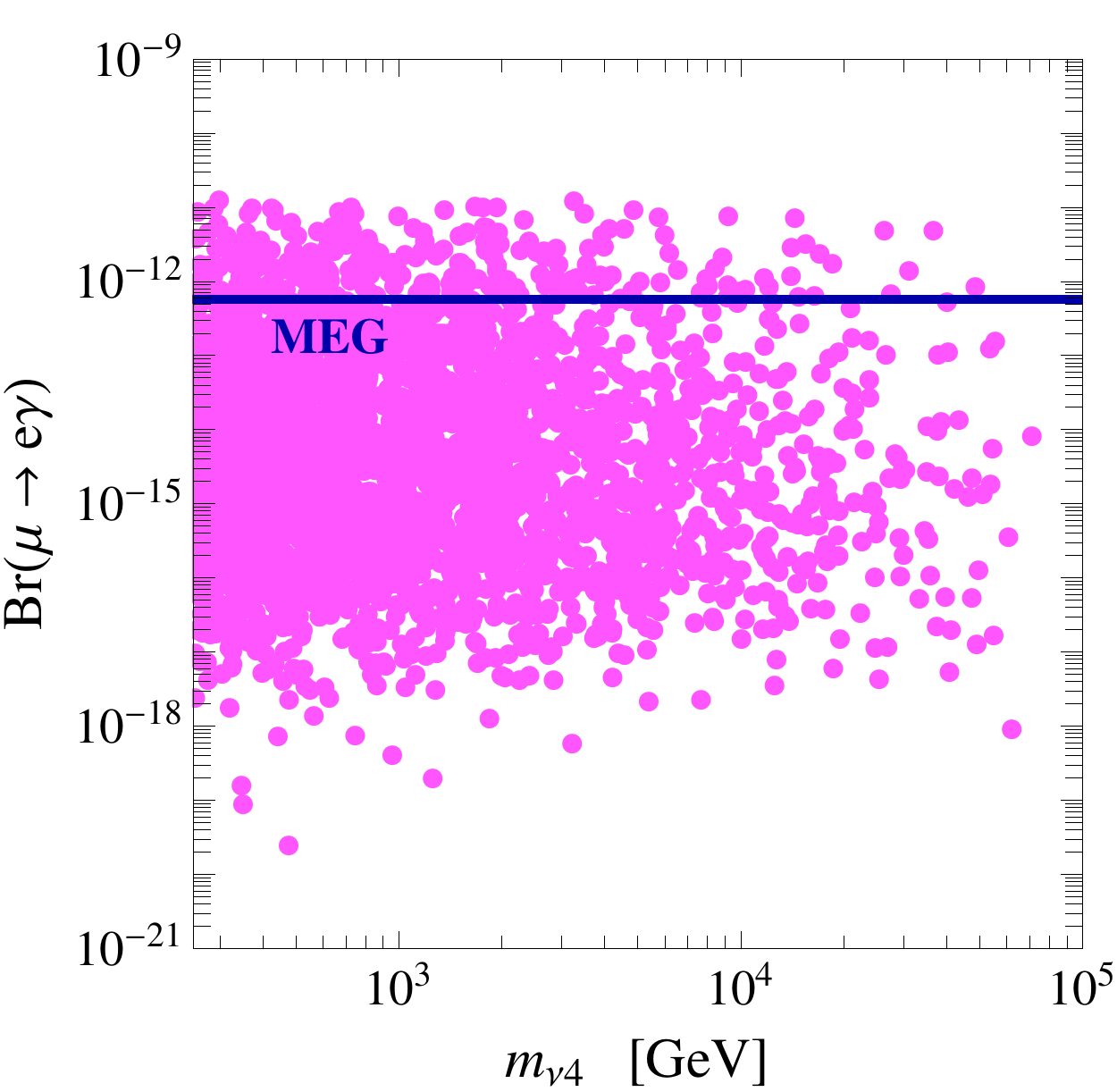}
\endminipage\hfill
\minipage{0.5\textwidth}
  \includegraphics[width=1.\linewidth]{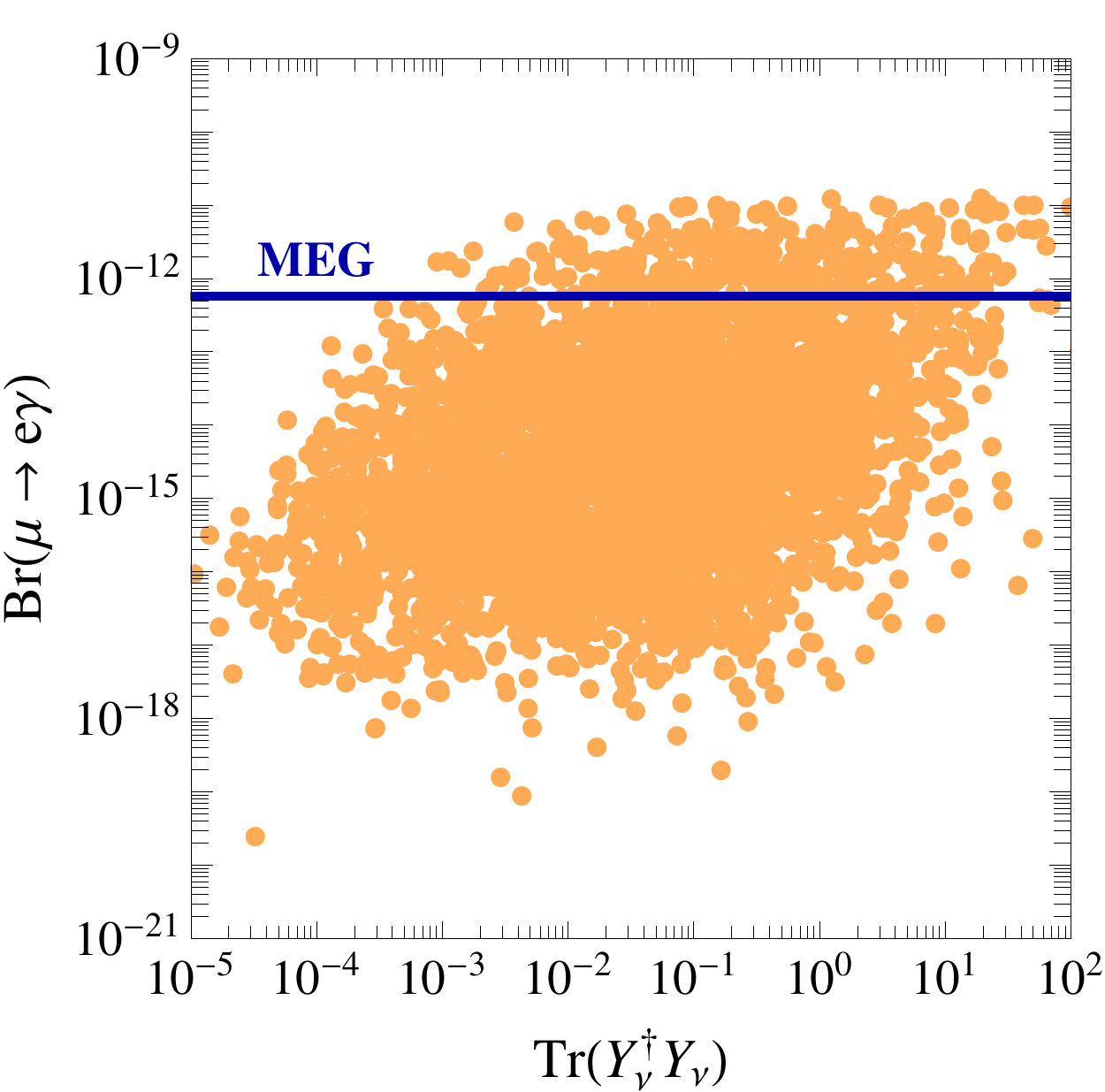}
\endminipage\\
\minipage{0.5\textwidth}
  \includegraphics[width=1.\linewidth]{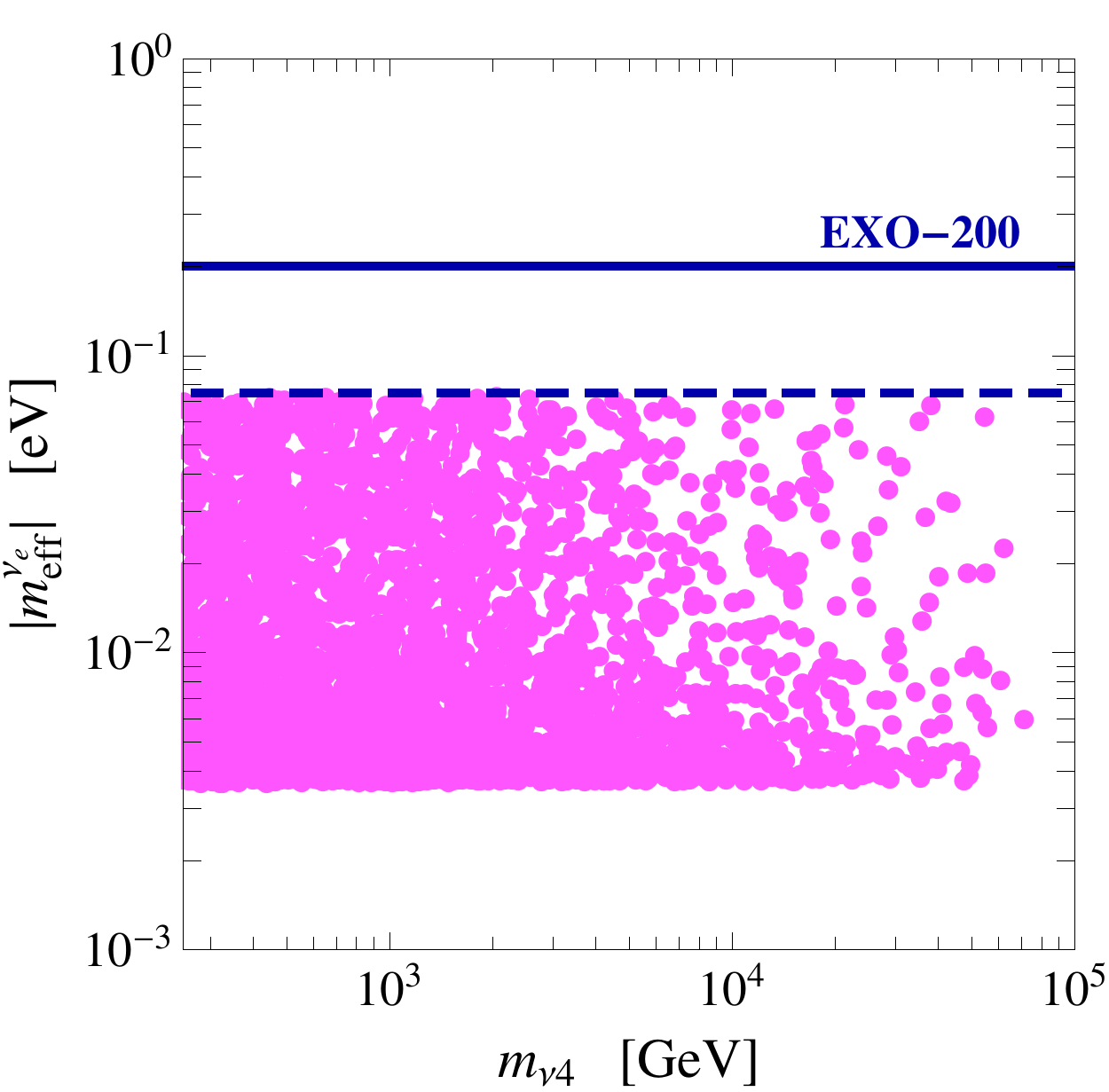}
\endminipage\hfill
\minipage{0.5\textwidth}
  \includegraphics[width=1.\linewidth]{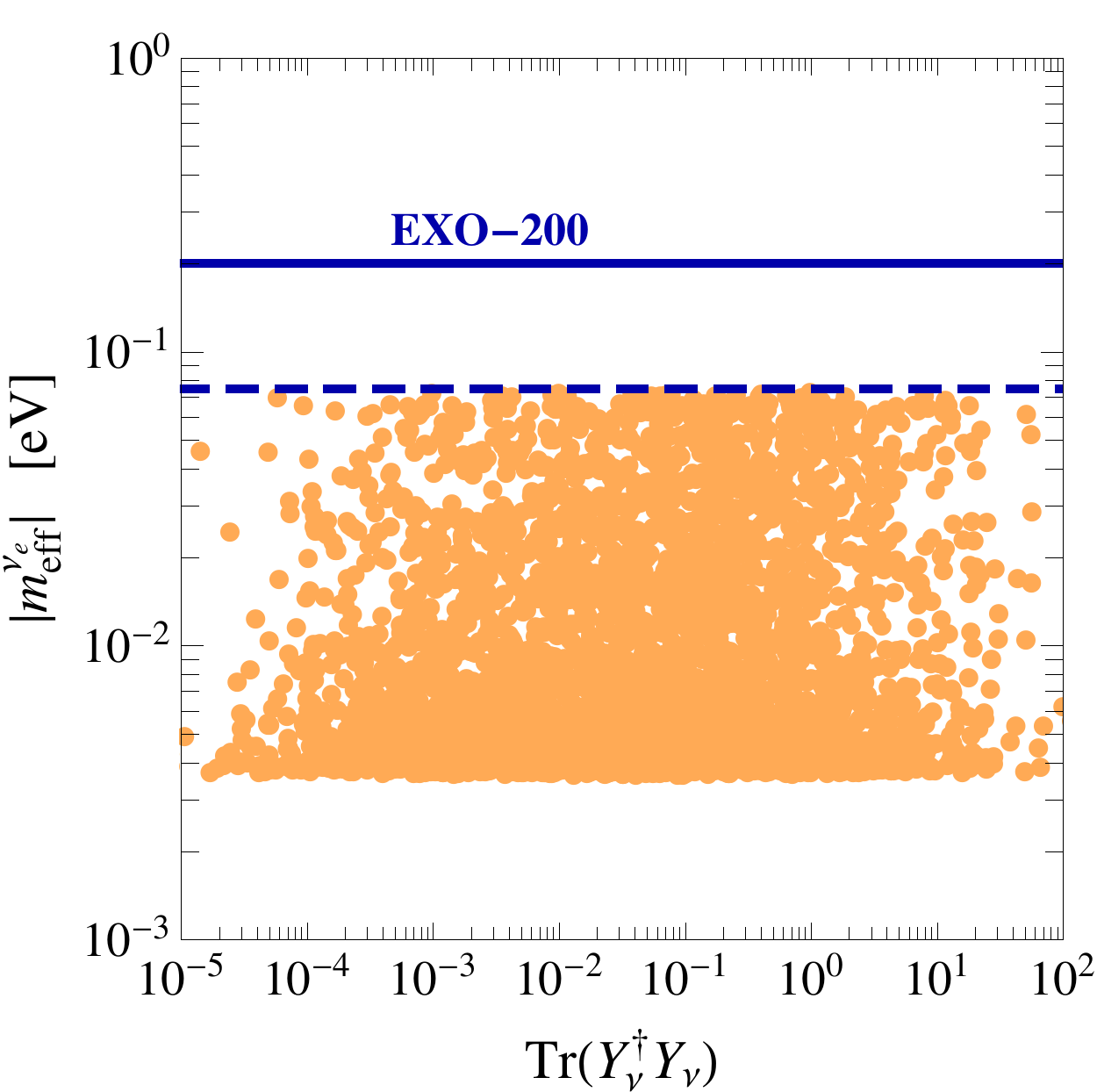}
\endminipage\\
\caption{\it Upper panel. Branching ratio for the decay process $\mu \to e\gamma$ as a function of the mass of the lightest sterile neutrino (left panel) and the 
trace
 of the Yukawa couplings ${\rm Tr}(Y^{\dag}_{\nu}Y_{\nu})$ (right panel). 
The blue horizontal line represents the upper bound set by the MEG collaboration~\cite{Adam:2013mnn}. 
Lower panel. Effective neutrino mass as a function of the mass of the lightest sterile neutrino (left panel) and the 
trace
 of the Yukawa couplings ${\rm Tr}(Y^{\dag}_{\nu}Y_{\nu})$ (right panel). The blue solid (dashed) line represents the upper bound (future sensitivity) of the EXO-200 experiment~\cite{Albert:2014awa}. Details about the numerical scan are given in the text.
All points comply with the bounds discussed in section~\ref{sec:BoundsLowEnergy}.
}\label{fig:FirstScan}
\end{figure}
We show our results in fig.~\ref{fig:FirstScan} for the normal ordering and in fig.~\ref{fig:FirstScanIO} for the inverted ordering.
In the upper (lower) panels of both figures we show the branching ratio ${\rm Br}(\mu \to e\gamma)$ (the effective neutrino mass  $m_{\rm eff}^{\nu_e}$)
as a function of the lightest heavy neutrino mass $m_{\nu 4}$ (plot on the left) and the trace
 of the Yukawa couplings ${\rm Tr}(Y^{\dag}_{\nu}Y_{\nu})$ (plot on the right). Few comments are in order.
 \begin{figure}[!htb!]
\begin{center}
\fbox{\footnotesize INVERTED ORDERING}
\end{center}
\vspace{-0.5cm}
\minipage{0.5\textwidth}
  \includegraphics[width=1.\linewidth]{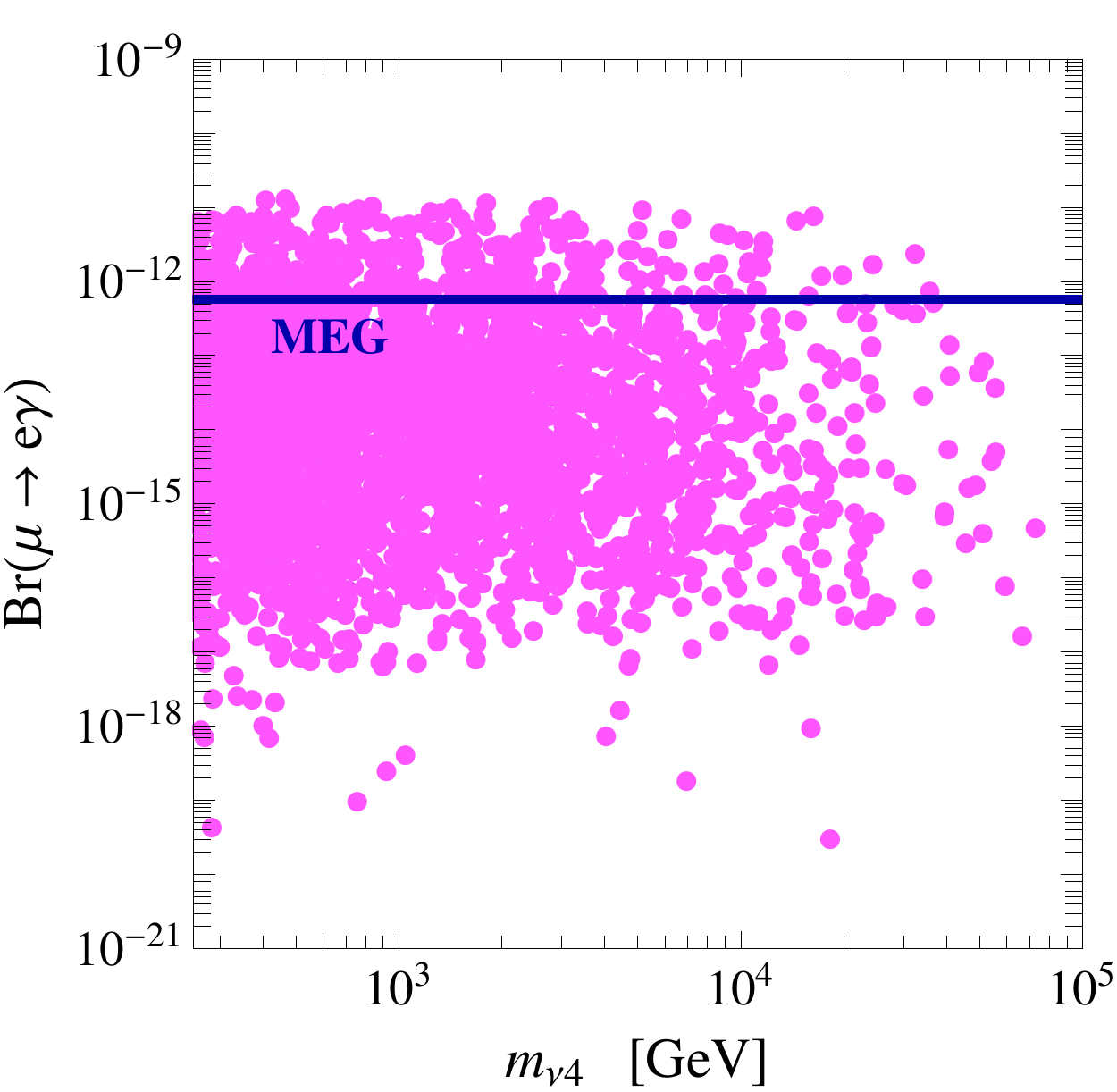}
\endminipage\hfill
\minipage{0.5\textwidth}
  \includegraphics[width=1.\linewidth]{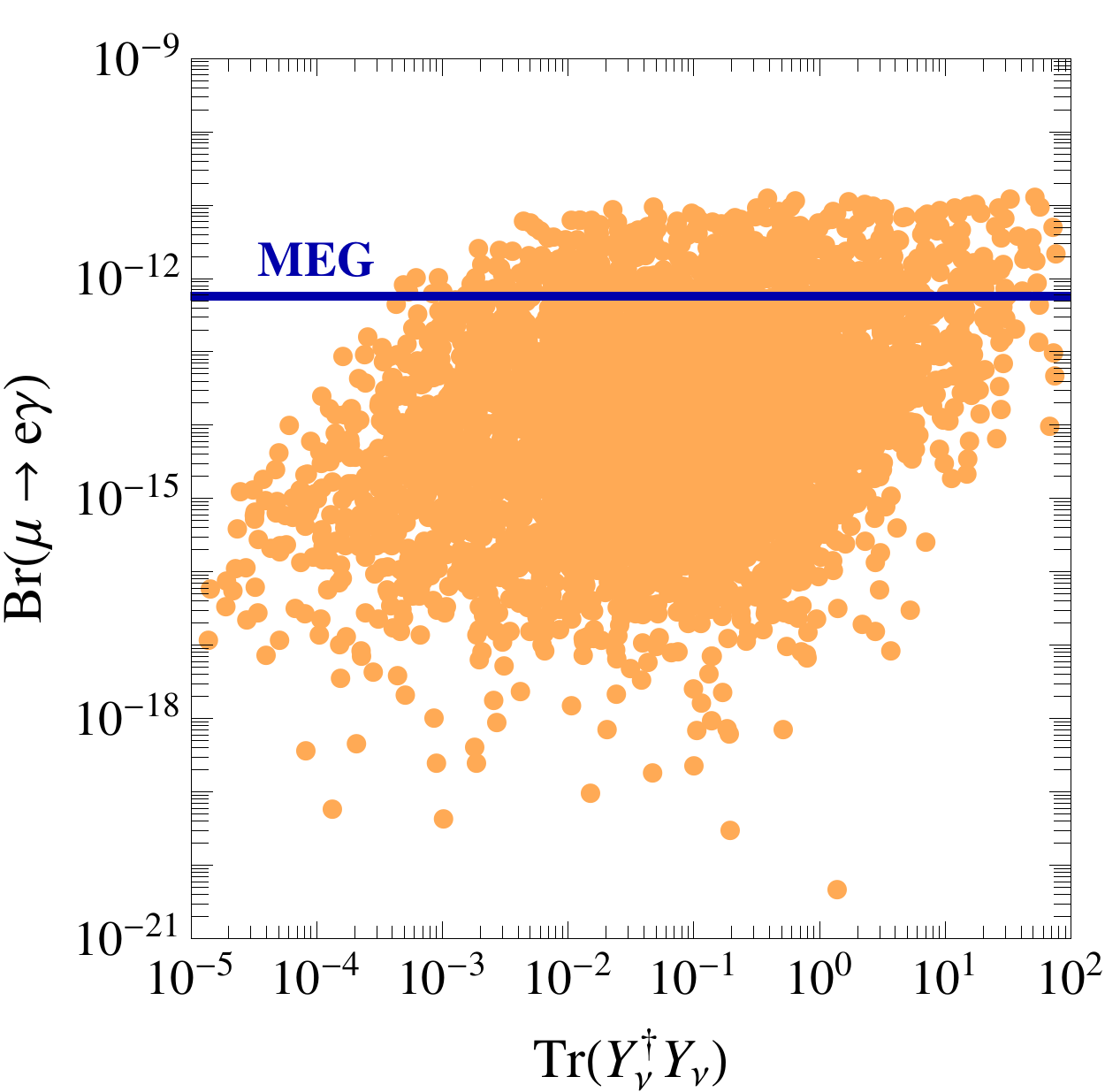}
\endminipage\\
\minipage{0.5\textwidth}
  \includegraphics[width=1.\linewidth]{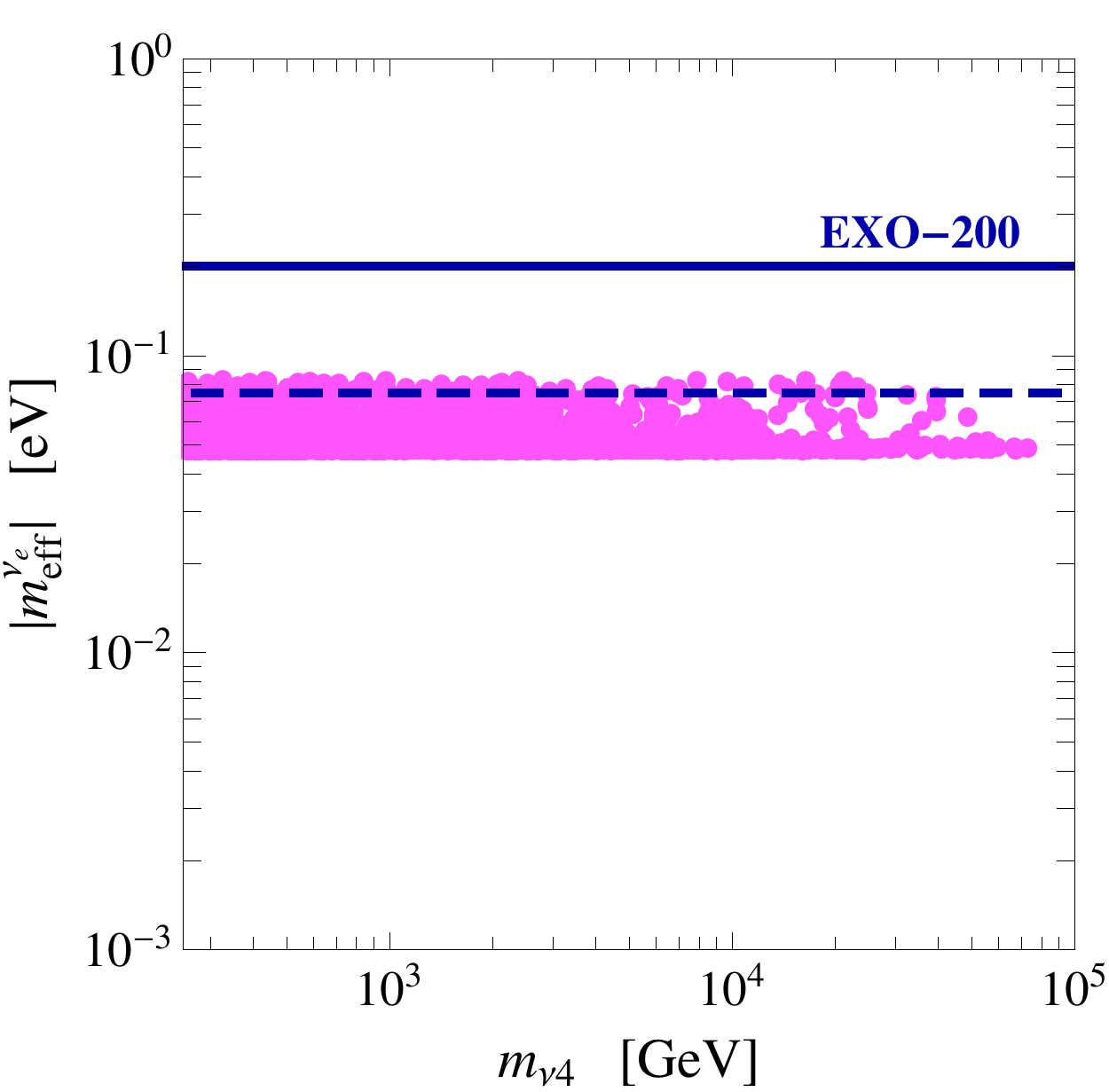}
\endminipage\hfill
\minipage{0.5\textwidth}
  \includegraphics[width=1.\linewidth]{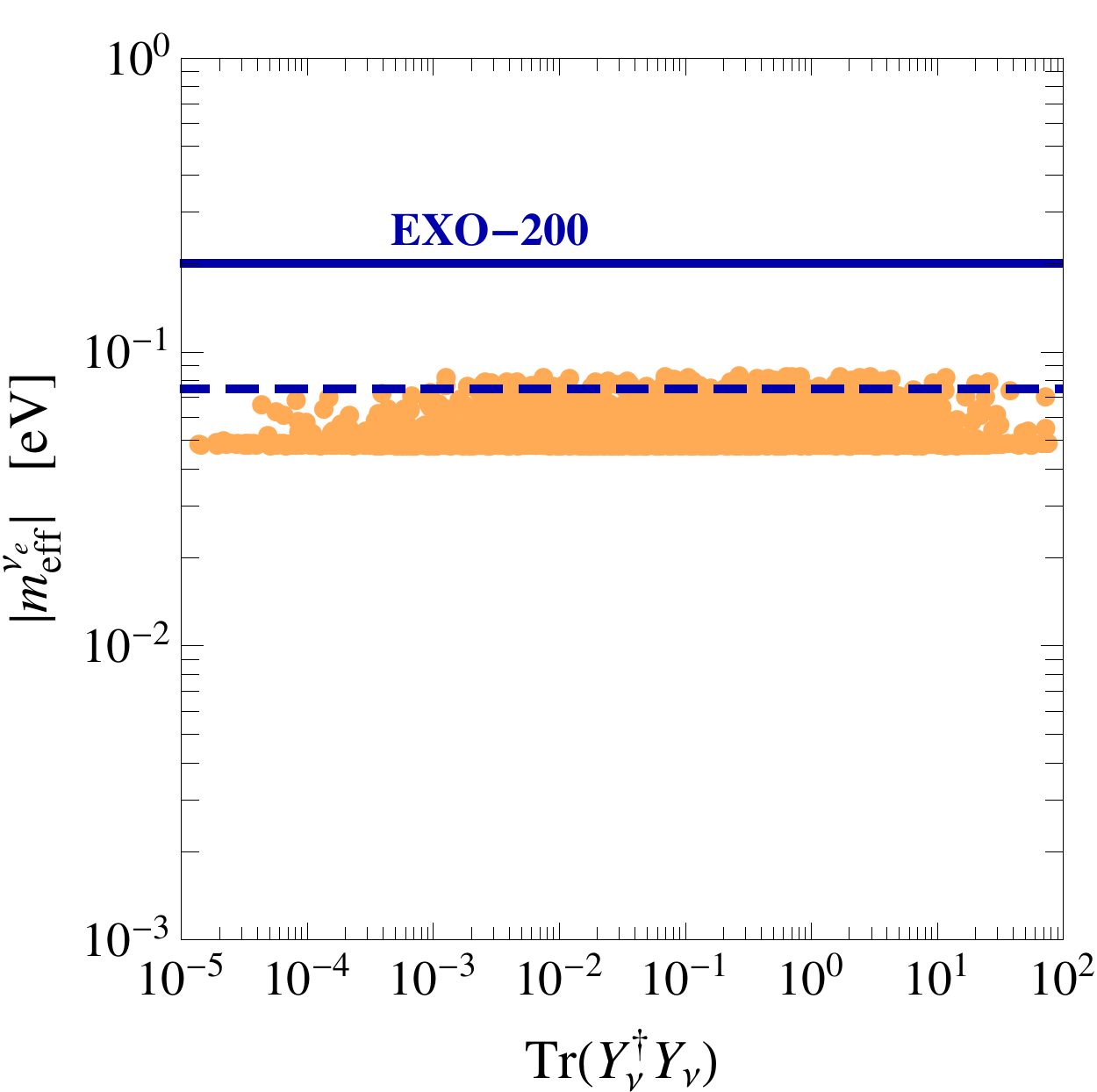}
\endminipage\\
\caption{\it Same as in fig.~\ref{fig:FirstScan}, but considering the inverted ordering.}\label{fig:FirstScanIO}
\end{figure}
\begin{enumerate}

\item  Normal and inverted ordering produce very similar distributions  considering the radiative decay ${\rm Br}(\mu \to e\gamma)$. This is caused by the well-known fact that the contribution of light active neutrinos is strongly suppressed by the extremely small value of light neutrino masses. 
In the ISS model a non-zero contribution to ${\rm Br}(\mu \to e\gamma)$ is entirely generated by the additional heavy neutrinos, and controlled 
by the mixings $U_{\mu i}, U_{e i} \sim m_D/M_R$ (see eq.~(\ref{eq:MixingMatrix}) in section~\ref{sec:Matching}).
We notice that in our scan we can obtain a signal close to the present experimental bound even considering $m_{\nu 4}$ as large as $10$ TeV 
and 
${\rm Tr}(Y^{\dag}_{\nu}Y_{\nu})$ as small as $10^{-3}$. For completeness we show in the left panel of fig.~\ref{fig:Temp} 
the result of our numerical scan in the plane $[{\rm Tr}(Y^{\dag}_{\nu}Y_{\nu}), m_{\nu 4}]$. We mark in dark 
cyan points with ${\rm Br}(\mu \to e\gamma) \geqslant 10^{-13}$. 
Points with large Yukawa couplings (e.g. ${\rm Tr}(Y^{\dag}_{\nu}Y_{\nu}) \gtrsim 0.5$) 
and sizable $\mu \to e\gamma$ rate are generated in the whole interval of analyzed masses for the right-handed neutrinos.

\item Normal and inverted ordering produce completely different distributions  considering the effective neutrino mass  $m_{\rm eff}^{\nu_e}$.
In this case the contributions of additional heavy neutrinos decouple since their masses are much larger than the typical momentum scale 
$p^2 \simeq -(125\,{\rm MeV})^2$. Therefore, in our numerical scan the ISS model resembles the typical scenario with only three light active neutrinos. 
The situation is well represented by the right panel of fig.~\ref{fig:Temp} where we show the effective neutrino
 mass for the normal and inverted ordering as a function of minimal neutrino mass. The normal ordering is suppressed since the largest neutrino mass 
 is multiplied by the small value of $s_{13}$.
However, in both cases the effective neutrino mass is close to the future sensitivity of the EXO-200 experiment.

\end{enumerate}

\subsubsection{Additional remarks}\label{sec:addRemarks}

Let us close this section summarizing further predictions and constraints on the inverse seesaw scenario.
The aim of the following discussion is to provide 
additional motivations enforcing the phenomenological relevance of EW-scale right-handed neutrinos with $\mathcal{O}(1)$ Yukawa couplings.

{\it Collider searches at the LHC}. At the LHC right-handed neutrinos with a mass 
not far above or below the Higgs mass and with a sizable Yukawa coupling $Y_{\nu} \gtrsim \mathcal{O}(10^{-2})$ 
affect the Higgs decay $h \to l\bar{l}\nu\bar{\nu}$ (see~\cite{BhupalDev:2012zg} for a recent analysis). 
Present bounds hold in the range $60 \leqslant M_R \leqslant 200$ GeV
with $10^{-2} \leqslant Y_{\nu} \leqslant 2$.\footnote{These bound were obtained in~\cite{BhupalDev:2012zg} considering a simplified setup with only one light flavor of heavy
neutrinos. Consequently, here $Y_{\nu}$ indicates the corresponding Yukawa coupling to the Higgs doublet.} 
Furthermore, the CMS collaboration placed upper limits on the active-sterile neutrino mixings 
in the same mass range for $M_R$ considering direct production of heavy neutrinos~\cite{Chatrchyan:2012fla,Atre:2009rg}.\footnote{See~\cite{Antusch:2015mia} for prospects at future lepton colliders.}
As stated in 
the introduction, these values of masses and couplings may have an impact on the stability of the EW vacuum, 
thus providing an additional motivation for the analysis that we shall perform in the next section. 

{\it Fit of LEP data via oblique parameters}. The fit of LEP data still provides today an important constraint on beyond the SM physics.
 The presence of additional sterile neutrinos modifies the oblique radiative corrections~\cite{Peskin:1990zt,Golden:1990ig,Holdom:1990tc,Peskin:1991sw,Altarelli:1990zd,Altarelli:1991fk,Barbieri:2003pr,Barbieri:2004qk}. In~\cite{Akhmedov:2013hec} it was shown that 
 right-handed neutrino masses of the order $M_{Ri}\sim \mathcal{O}(10)$ TeV, together with violation of unitarity of the order $\epsilon_{\alpha\beta} \sim 10^{-3}\div 10^{-6}$, can improve the fit of LEP data with respect to the SM.

{\it Leptogenesis}. In the inverse seesaw scenario the decay of (nearly degenerate) heavy Majorana neutrinos  
can realize the so-called resonant leptogenesis~\cite{Pilaftsis:2003gt}. Remarkably, resonant leptogenesis
can be realized with heavy Majorana neutrinos as light as $1$ TeV~\cite{Gu:2010xc} (in contrast with the usual 
thermal leptogenesis, realized in the type-I seesaw, in which $M_R \gtrsim 10^9$ GeV~\cite{Sacha}).

{\it Naturalness}. On a general ground, whenever a threshold with particles of mass $M$ coupled to the Higgs with strength $\xi$ is present,
quantum corrections generate a contribution $\delta m_H^2 \approx \xi^2 M^2/16\pi^2$ to the renormalized Higgs boson mass. 
If $\delta m_H^2 \gg v^2$, 
an unnatural cancellation between $\delta m_H^2$ and the bare Higgs mass is required in order to reproduce the observed value of the Higgs boson mass.
The condition $\delta m_H^2  \lesssim v^2$ can be used as a criterium to construct natural model of new physics~\cite{Farina:2013mla,Fabbrichesi:2015zna}.
In the context of inverse seesaw models, the scale $M$ is the mass of right-handed heavy neutrinos $M_R$, and the coupling $\xi$ is the Yukawa coupling $Y_{\nu}$.
TeV-scale values of $M_R$ and $\mathcal{O}(1)$ Yukawa couplings satisfy the naturalness condition.

To sum up, the presence of heavy neutrinos with a mass not far above the EW scale and a sizable Yukawa coupling with the Higgs boson 
has extremely rich phenomenological consequences.
Motivated by the results achieved in this section, we are now in the position to tackle the second part of our analysis in which we aim to investigate the impact of heavy neutrinos on the stability of the EW vacuum.

\section{Stability of the electroweak vacuum and low-scale seesaw}\label{sec:Stability}

The key ingredient in the study of the stability of the EW potential is represented by the quantum effective potential $V_{\rm eff}$.
In particular, we search for the instability scale $\Lambda$ corresponding to the Higgs field value for which the potential becomes smaller than its value at the EW minimum. If such scale does not exist the vacuum is called \emph{absolute stable}, otherwise the Higgs vacuum is not the global minimum, and a quantum tunnelling to the true vacuum may occur. The decay probability can be computed from the bounce solution of the euclidean equations of motion of the Higgs field~\cite{Coleman:1977py,Callan:1977pt}. If the corresponding lifetime is much bigger that the age of the Universe the EW vacuum is called \emph{metastable}. 
This can be translated into a lower bound on the Higgs effective quartic coupling $\lambda_{\rm eff}$, driven to negative values, that at leading order reads as
\begin{equation}\label{eq:MasterBound}
\left|\lambda_{\rm eff}(\mu)\right| >  \frac{8 \pi^2}{3} \frac{1}{\log (\tau \mu)}~,
\end{equation} 
where $\tau$ is the age of the Universe $\tau = 4.35 \times 10^{17}$ sec~\cite{Ade:2013sjv} and $\mu$ is the renormalization scale of the RG running. The Higgs effective quartic coupling will be defined in the next section.  The SM quantum corrections at zero temperature to the metastable condition have been computed in~\cite{Isidori:2001bm} but are not considered in this work. The measured values of the Higgs and the top masses place the SM in a metastable position in the phase space diagram~\cite{Buttazzo:2013uya}. The inclusion of extra fermionic degrees of freedom, as right-handed neutrinos in seesaw extensions, can clearly change this picture possibly driving the model to an instability region. 
Therefore, by requiring the metastability of the EW vacuum, together with the perturbativity of the gauge, scalar and Yukawa couplings up to the Planck mass, we can constrain the parameter space of low-scale seesaw models.
 
In the following sections we will describe the theoretical tools needed in our stability analysis: the Higgs effective quartic couplings which determines the instability scale, the RGEs in the $\overline {\rm MS}$ describing the running of the couplings from the EW scale up to the Planck scale and, finally, the matching conditions which provide the initial values of the $\overline {\rm MS}$ parameters from the EW physical observables. 

\subsection{The Higgs effective quartic coupling}

From a preliminary analysis one can show that the vacuum instability appears at a scale much bigger than the EW minimum. This allows us to consider the effective potential at large field values, $\phi \gg v$, and to use the following approximation
\begin{equation}\label{eq:EffPot}
V_{\rm eff}(\phi, t)\approx \frac{\lambda_{\rm eff}(\phi, t)}{4}\phi^4~,
\end{equation}
where the Higgs quadratic mass term has been neglected and $t$ is the logarithm of the renormalization running scale $\mu$. The effective quartic coupling $\lambda_{\rm eff}$ is extracted from the RG-improved effective potential at two-loop order in the SM~\cite{Ford:1992pn} and at one-loop for the right-handed neutrino corrections, computed in the $\overline{\rm MS}$ renormalization scheme and in the Landau gauge.  The complete two-loop expression is too lengthy to be given here. We show, instead, the effective quartic coupling at one-loop order in the SM~\cite{Casas:1994qy}
\begin{equation}\label{eq:LambdaEffSm}
\lambda_{\rm eff}(\phi, t) \approx e^{4\Gamma(t)}\left\{
\lambda(t) + \frac{1}{16\pi^2}\sum_{p=W,Z,h,\chi,t}N_p  \kappa_p^2(t)\left[
\ln\frac{\kappa_p(t)e^{2\Gamma(t)}\phi^2}{\mu(t)^2}
- C_p \right]
\right\}~,
\end{equation}
where the $p$-coefficients are summarized in table~\ref{tab:pcoeff} and $\Gamma(t)$ is defined as
\begin{equation}
\Gamma(t) = \int_0^t d t' \gamma(t')~,
\end{equation}
with $\gamma$  the anomalous dimension of the Higgs field.
\begin{table}[!htb!]
\begin{center}
\begin{tabular}{|c||c|c|c|c|c|c|c|}
\hline
  & $t$ & $W^{\pm}$ & $Z$ & $h$  & $\chi^{\pm}$  &  $\chi^0$ \\ \hline\hline
 $N_p$  & -12 & 6 & 3 & 1 & 2 & 1  \\ \hline
  $C_p$  & 3/2 & 5/6 & 5/6 & 3/2 & 3/2 & 3/2   \\ \hline
    $\kappa_p$  & $y_t^2/2$ & $g_2^2/4$ & $(g_2^2 + g_Y^{2})/4$ & $3\lambda$ & $\lambda$ & $\lambda$   \\ \hline
\end{tabular}
\end{center}
\caption{\it SM p-coefficients entering in eq.~(\ref{eq:LambdaEffSm}).}\label{tab:pcoeff}
\end{table}

The contribution of the heavy neutrinos to the RG-improved effective potential is~\cite{Quiros:1999jp}
\begin{equation}\label{eq:EffPot}
\Delta V_{\rm eff}^{\nu}(\phi,t)
=
-\frac{1}{32\pi^2}\sum_{i=1,2,3}
\theta_{\nu i}(t)\,2\mathcal{M}_{\nu i}^{4}(\phi, t)\left[\ln\frac{\mathcal{M}_{\nu i}^{2}(\phi,t)}{\mu(t)^2} - \frac{3}{2} \right]~,
\end{equation}
where $\theta_{\nu i}(t) = \theta(t - \ln M_{R i}/\mu_0)$ and $\mathcal{M}_{\nu i}(\phi,t)$ are the three (double degenerate) non-zero eigenvalues of the $9\times 9$ mass matrix
\begin{equation}\label{eq:MassMatrixRGE}
\mathcal{M}_{\phi}(t) =
\left(
\begin{array}{ccc}
 0  & Y_{\nu}(t)^T\phi(t)/\sqrt{2}  & 0  \\
Y_{\nu}(t)\phi(t)/\sqrt{2} &  0 & \hat{M}_{Ri}     \\
  0 & \hat{M}_{Ri} & 0   \\
\end{array}
\right)~,
\end{equation}
with $\hat{M}_{Ri} \equiv {\rm diag}(M_{R1},M_{R2},M_{R3})$ and $\phi(t) = e^{\Gamma(t)}\phi$. Compared with eq.~(\ref{eq:MasterMassMatrix}), we 
are considering the $\mu_{\rm S}\to 0$ limit in which the three light neutrinos are massless.

The factor of two in eq.~(\ref{eq:EffPot}) comes from the fact that each non-zero eigenvalue is double degenerate.

The RG running parameter $t$ can be chosen in such a way that the convergence of perturbation theory -- otherwise spoiled by the presence of large logs -- is improved.
We follow the prescription $\mu(t) = \phi$ widely used in vacuum stability analyses.

The contribution of heavy neutrinos to the effective potential
produces two distinctive effects on the RG-evolution of the effective quartic coupling in eq.~(\ref{eq:LambdaEffSm}).

\begin{itemize}

\item[$\circ$] Above each threshold $M_{Ri}$ and at high values of renormalization scale $\mu(t) \gg M_{Ri}$ it contributes explicitly to $\lambda_{\rm eff}$ 
introducing the correction
\begin{equation}\label{eq:NeutrinosEffPot}
\lambda_{\rm eff}^{\nu i}(\phi,t) = -\frac{1}{16\pi^2} 4\kappa_{\nu i}^4(t)
\left[\ln\frac{\kappa_{\nu i}^{2}(t)e^{2\Gamma(t)}\phi^2}{\mu(t)^2} - \frac{3}{2} \right]~,
\end{equation}
where $\kappa_{\nu i}(t)$ is the coefficient of the $\phi$-dependent part of $\mathcal{M}_{\nu i}(\phi,t)$.

\item[$\circ$] Below each threshold $M_{Ri}$ the corresponding heavy neutrinos are integrated out. The matching produces the threshold correction to the effective potential
\begin{equation} \label{eq:DeltaVeff}
\Delta V_{\rm th}^{\nu i}(\phi, t)
=
-\frac{1}{32\pi^2}
2\mathcal{M}_{\nu i}^{4}(\phi, t)\left[\ln\frac{\mathcal{M}_{\nu i}^{2}(\phi,t)}{M_{Ri}^2} - \frac{3}{2} \right]~,
\end{equation}
which translates into a threshold correction $\Delta \lambda_{\rm th}^{\nu i}$ to the Higgs quartic coupling at the $M_{Ri}$ mass scale which can be extracted from the $\phi^4$ term of eq.~\ref{eq:DeltaVeff} and it is explicitly given by
\begin{equation}
\Delta \lambda_{\rm th}^{\nu i}(t) = \frac{1}{6}\left. \frac{d \Delta V_{\rm th}^{\nu i}(\phi, t)}{d\phi}\right|_{\phi = 0}~.
\end{equation}

\end{itemize}

\subsection{The matching conditions}\label{sec:Matching}

The RG equations employed in this work are computed in the $\overline {\rm MS}$ renormalization scheme and must be equipped with suitable initial conditions for the running parameters evaluated in the same scheme. These parameters are expressed in terms of physical observables, defined in the on-shell OS scheme, through appropriate relations called matching conditions. For a generic parameter $\alpha$ the matching at the scale $\mu$ is given by
\bea
\alpha(\mu) = \alpha_{\rm OS} - \delta \alpha_{\rm OS} + \delta \alpha_{\overline {\rm MS}}~,
\eea
where $\alpha(\mu)$ and $\alpha_{\rm OS}$ are respectively the $\overline {\rm MS}$ and the OS parameters, while $\delta \alpha_{\overline {\rm MS}}$ and $\delta \alpha_{\rm OS}$ are the corresponding counterterms in the two schemes. The difference between the two of them is an ultraviolet-finite correction.

The SM parameters obtained from the matching procedure are $(\lambda, y_t, g_1, g_2)$, while the QCD coupling constant $g_3$ is directly extracted from the $\alpha_3(M_Z)$ which is already defined in the $\overline {\rm MS}$ scheme. The physical observables used to compute the input values of the $\overline {\rm MS}$ parameters are given in table~\ref{tab:inputvalues}. \\
\begin{table}[!htb!]
\begin{center}
\small
\begin{tabular}{|c|c|c|}\hline
\textbf{Name} & \textbf{Value} & \textbf{Description} \\
\hline\hline
$M_W$ & 80.384 GeV & $W$ boson pole mass \\
$M_Z$  & 91.1876 GeV & $Z$ boson pole mass \\
$M_h$  & 125.09 GeV & Higgs boson pole mass \\
$M_t$  & 173.34 GeV & Top quark pole mass \\
$v \equiv (\sqrt{2} G_\mu)^{-1/2}$ & 246.21971 GeV & Higgs vev from the $\mu$ decay \\
$\alpha_3(M_Z)$ & 0.1184 & $\overline {\rm MS}$ QCD structure constant (5 flavors) \\
\hline
\end{tabular}
\end{center}
\caption{\it Physical observables used to extract the SM parameters in the $\overline {\rm MS}$ scheme through the matching procedure. For the Higgs mass we used the latest result \cite{Aad:2015zhl}, for all the other parameters we refer to \cite{Buttazzo:2013uya}.}\label{tab:inputvalues}
\end{table}
The details of the strategy can be found in \cite{Buttazzo:2013uya} where the SM two-loop (NNLO) corrections to the matching conditions have been discussed. In particular, in \cite{Buttazzo:2013uya} a complete two-loop analysis has been performed in the EW sector and the N3LO (three-loop) pure QCD effect has been included in the matching of the top Yukawa coupling and the strong coupling constant. The running of the latter from $M_Z$ to $M_t$, which is the starting scale of our stability analysis, has been performed including the QCD four-loop $\beta$ function. 

In low-scale seesaw extensions, the right-handed neutrino corrections to the matching conditions can be important and must be taken into account. In our analysis we have considered all the SM results given in \cite{Buttazzo:2013uya}, supplemented by new physics contributions computed at one-loop order from the Lagrangian in eq.~(\ref{eq:LISS}). 
In particular, the additional neutrinos introduce corrections to the masses of the gauge bosons $M_Z$, $M_W$, the Higgs $M_h$, the quark top $M_t$ and to the muon decay, from which the Fermi constant $G_\mu$ is extracted. These corrections depend on the masses of the heavy neutrinos, their interactions with the SM fields mediated by the Yukawa couplings $Y_\nu$ (obtained from the parameterization in eq.~(\ref{eq:CasasIbarra})), and the full $9\times9$ mixing matrix $U$. Due to the mass hierarchy $M_R \gg m_D$, and to the smallness of the light neutrino masses -- which can be safely neglected in the computation of the matching conditions at the EW scale -- the mixing matrix $U$ can be expanded in the ratio $m_D/M_R$, with $\mu_S \rightarrow 0$, as
\begin{equation}\label{eq:MixingMatrix}
U= \left(
\begin{array}{ccc}
1 & \frac{1}{\sqrt{2}} m_D^\dag M_R^{-1} & \frac{i}{\sqrt{2}} m_D^\dag M_R^{-1} \\
0 & \frac{1}{\sqrt{2}} & - \frac{i}{\sqrt{2}} \\
- M_R^{-1} m_D & \frac{1}{\sqrt{2}} & \frac{i}{\sqrt{2}}
\end{array}
\right) + \mathcal O\left( \frac{m_D^2}{M_R^2}\right) ~.
\end{equation}
Notice that the PMNS block has been set to the unit matrix, consistently with the approximation $m_{\nu i} \simeq 0$.   

Moreover, we have verified the Appelquist-Carazzone theorem in the new physics sector. In particular, we have checked the decoupling of right-handed neutrino contributions from the matching conditions in the limit of large Majorana masses. This has to be expected since these masses are not generated by spontaneous symmetry breaking of the Higgs field.

Concerning the matching conditions of the Yukawa matrix $Y_\nu$, 
computing perturbative corrections to the matching conditions at the EW scale will not considerably improve the precision on the determination of $Y_\nu$ in the $\overline {\rm MS}$ scheme
(since in any case $Y_\nu$ turns out to be related to unknown parameters by means of eq.~(\ref{eq:CasasIbarra})). 
Therefore, for the sake of simplicity, we decided to match the $\overline {\rm MS}$ Yukawa matrix $Y_\nu$ to its OS version at the tree-level, namely $Y_\nu(M_t) \simeq Y_\nu$.

\subsection{The renormalization group equations}\label{sec:Beta}

All the dimensionless couplings $(\lambda, y_t, g_i,Y_{\nu})$ are evolved from the top-mass scale, $M_t$, up to the Planck scale using the three-loop (NNLO) RG equations for the SM parameters and the two-loop (NLO) $\beta$ functions for the Yukawa matrix $Y_\nu$. Here $g_i$ stands for the three gauge coupling constants and we have retained only the top-quark contribution in the SM Yukawa sector. The system of coupled RG equations is then given by 
\begin{eqnarray} 
\frac{d\lambda(t)}{dt}&=& \beta_{\lambda}(\lambda, y_t, g_i,Y_{\nu})~,\label{eq:RGsystem1}\\
\frac{d y_t(t)}{dt}&=& \beta_{y_t}(\lambda, y_t, g_i,Y_{\nu})~,\label{eq:RGsystem2}\\
\frac{d g_i(t)}{dt}&=& \beta_{g_i}(\lambda, y_t, g_i,Y_{\nu})~,\label{eq:RGsystem3}\\
\frac{d Y_{\nu}(t)}{dt}&=& \beta_{Y_{\nu}}(\lambda, y_t, g_i,Y_{\nu})~,\label{eq:RGsystem4}
\end{eqnarray}
where the $\beta$ functions are computed in perturbation theory in the $\overline{\rm MS}$ renormalization scheme. Due to their lengthy expressions, we present only the one-loop corrections to the r.h.s. of eqs.~(\ref{eq:RGsystem1}--\ref{eq:RGsystem4}), namely
\begin{eqnarray}
\beta_{\lambda} &=& \kappa \left[ 24 \lambda^2 + \lambda \left( 12 y_t^2 + 4 {\rm Tr} (Y_\nu^\dag Y_\nu) - \frac{9}{5} g_1^2 - 9 g_2^2\right) - 6 y_t^4 - 2 {\rm Tr}(Y_\nu^\dag Y_\nu)^2  \right. \nonumber \\
&+&\left. \frac{27}{200} g_1^4 + \frac{9}{8} g_2^4 + \frac{9}{20} g_1^2 g_2^2 \right]~, \label{eq:betalambda} \\
\beta_{y_t} &=& \kappa \left[ - \frac{17}{20} g_1^2 - \frac{9}{4} g_2^2 - 8 g_3^2 + \frac{3}{2} y_t^2 + {\rm Tr}(Y_\nu^\dag Y_\nu)   \right] y_t~, \label{eq:betayt} \\
\beta_{g_1} &=& \kappa \frac{41}{10} g_1^3~, \qquad  \beta_{g_2} = - \kappa \frac{19}{6} g_2^3~, \qquad \beta_{g_3} = - \kappa \, 7 g_3^3~, \\
\beta_{Y_{\nu}} &=& \kappa \left[ \left( -\frac{9}{20} g_1^2 - \frac{9}{4} g_2^2 + 3 y_t^2 + {\rm Tr}(Y_\nu^\dag Y_\nu) \right)  Y_{\nu} + \frac{3}{2} Y_\nu Y_\nu^\dag Y_\nu \right]~,\label{eq:betaynu}
\end{eqnarray}
where $\kappa = 1/(16\pi^2)$ and the abelian gauge coupling is given in the GUT normalization $g_1 = \sqrt{5/3} g_Y$. Notice that the RG equations given above are defined for a renormalization scale $\mu$ bigger than any particle mass of the model. For lower scales, heavy degrees of freedom are integrated out and the corresponding coupling must be removed by hand from the $\beta$ functions. Indeed, in the $\overline {\rm MS}$ renormalization scheme the decoupling of heavy degrees of freedom is not automatic and has to be explicitly implemented at the different particle thresholds.  

\begin{figure}[!htb!]
\minipage{0.5\textwidth}
  \includegraphics[width=.97\linewidth]{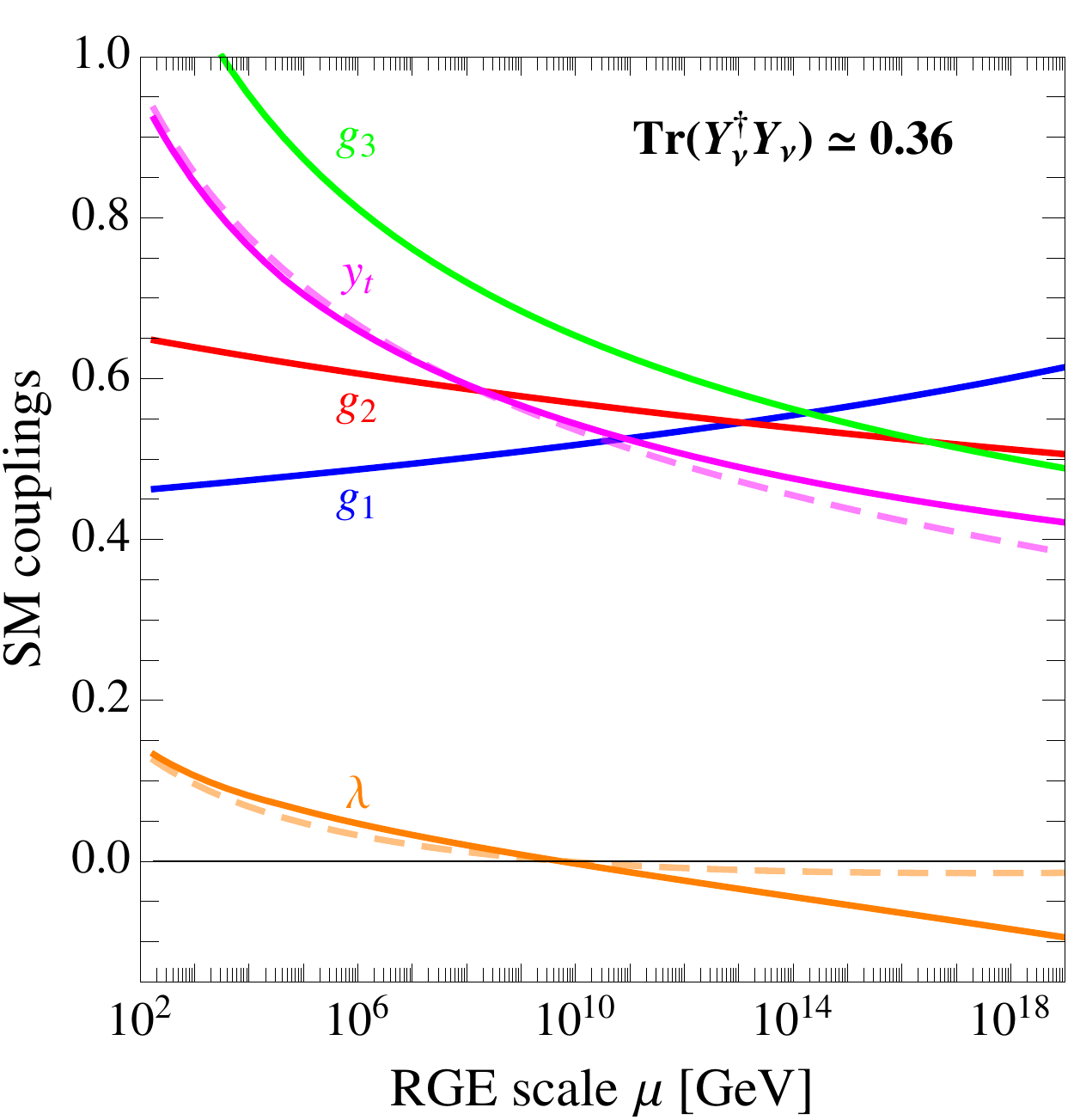}
\endminipage\hfill
\minipage{0.5\textwidth}
  \includegraphics[width=.97\linewidth]{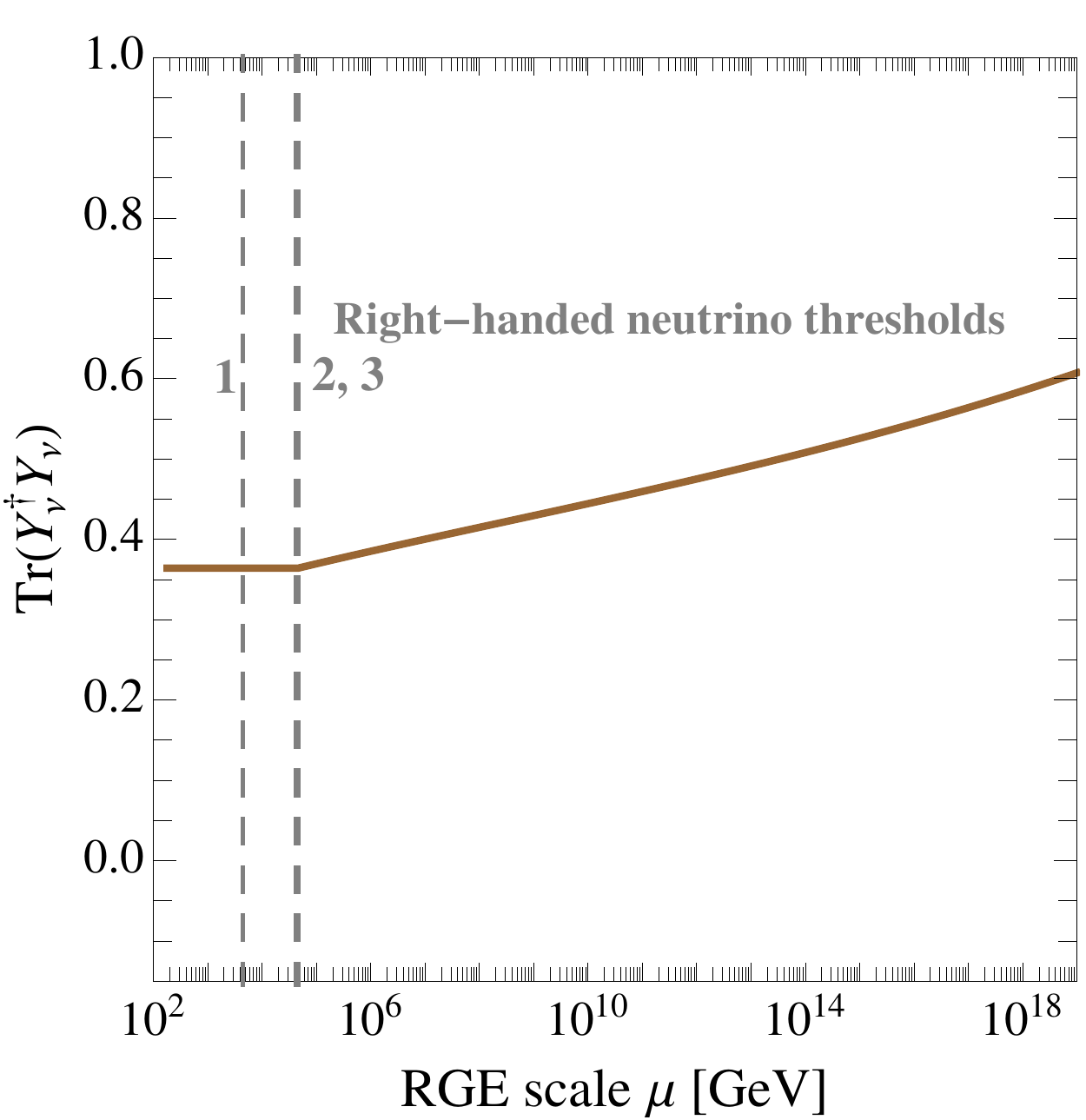}
\endminipage \vspace{.3 cm}
\caption{\it Left panel. RG evolution of the SM couplings. Solid lines take into account the effects of right-handed neutrinos in the inverse seesaw model with ${\rm Tr}(Y_\nu^\dag Y_\nu) \simeq 0.36$ at the EW scale. Dashed lines represent the running of couplings in the SM. Right panel. Evolution of ${\rm Tr}(Y_\nu^\dag Y_\nu)$. The heavy right-handed neutrino thresholds are explicitly shown.}\label{fig:RGE}
\end{figure}

\begin{figure}[!htb!]
\minipage{0.5\textwidth}
  \includegraphics[width=.97\linewidth]{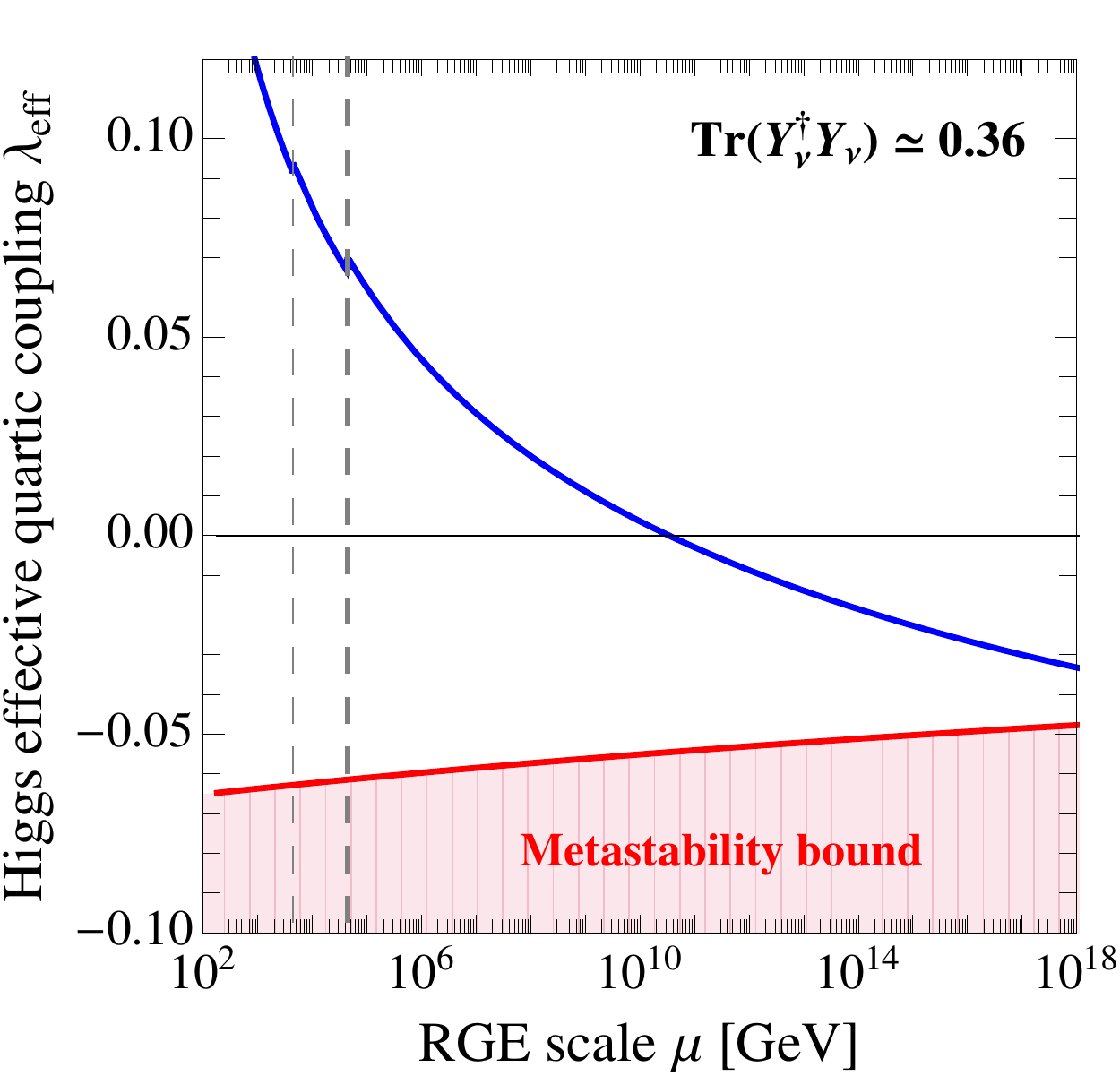}
\endminipage\hfill
\minipage{0.5\textwidth}
  \includegraphics[width=.97\linewidth]{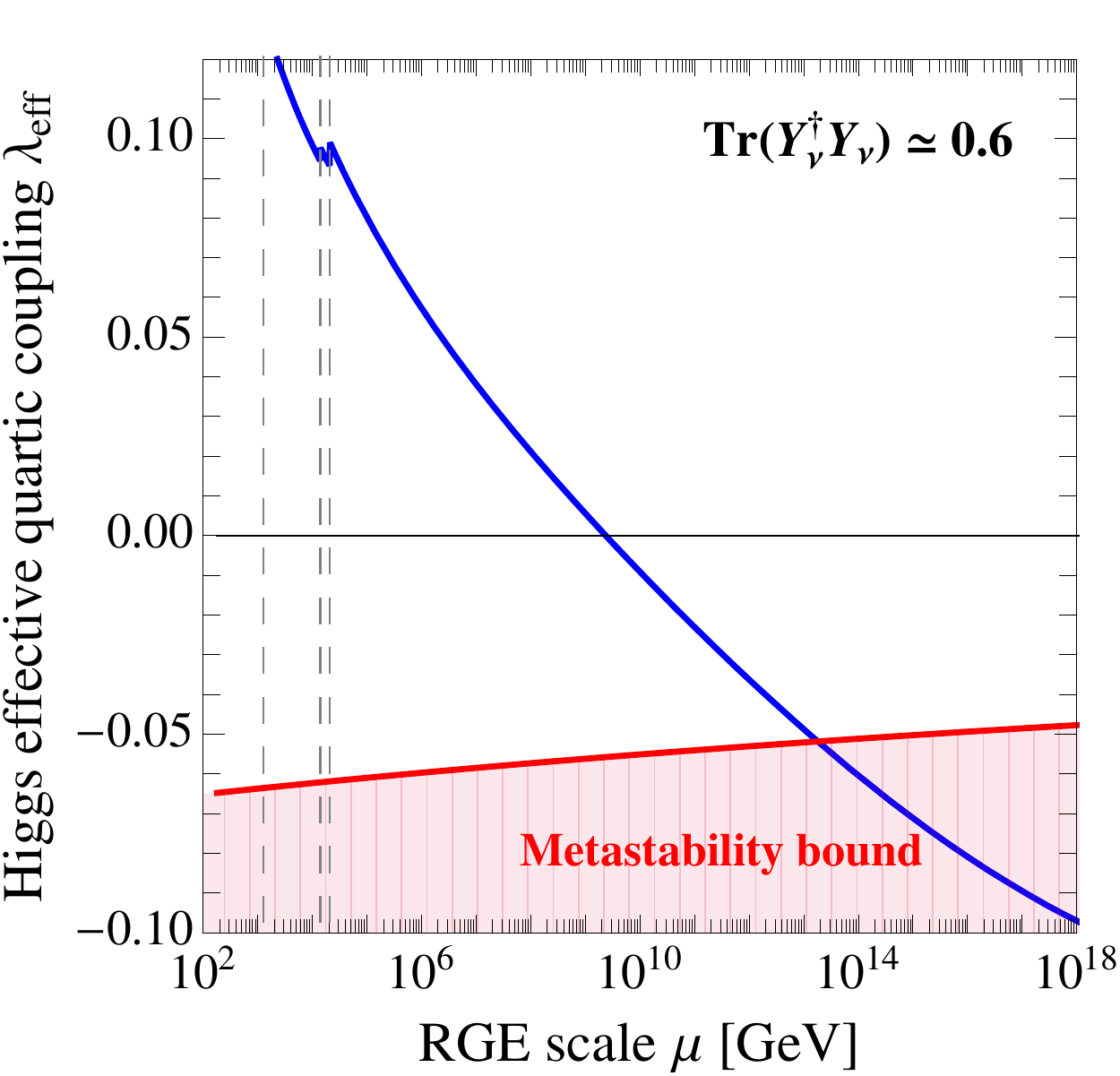}
\endminipage \vspace{.3 cm}
\caption{\it 
RG evolution of the effective Higgs quartic coupling in the inverse seesaw model in two different setup: ${\rm Tr}(Y_\nu^\dag Y_\nu) \simeq 0.36$ (left panel) and ${\rm Tr}(Y_\nu^\dag Y_\nu) \simeq 0.6$ (right panel). In the latter case the Yukawa couplings have a sizable impact on $\lambda_{\rm eff}$ and the metastability bound is violated below the Planck scale. }\label{fig:BoundMeta}
\end{figure}

As far as the vacuum stability is concerned, the right-handed neutrinos behave like the top-quark and drive $\lambda$ to negative values faster than the SM case.
This is clear from eq.~(\ref{eq:betalambda}). On the other hand, their impact on the top Yukawa coupling -- see eq.~(\ref{eq:betayt}) -- is to increase $y_t$ all along the RG evolution. Actually, this is another source of vacuum destabilization with respect to the SM picture, because a bigger value of $y_t$ has a bigger decreasing effect on the Higgs quartic coupling. Nevertheless, the overall behavior of the top-quark Yukawa coupling is dictated by the large and negative QCD corrections which lead to a decreasing $y_t$. These features can be easily deduced from fig.~\ref{fig:RGE}, left panel, where the running of the SM couplings is depicted. Here solid lines represent the evolution in the seesaw extended scenario, while dashed curves correspond to the pure SM. \\
Contrary to the top Yukawa case, the QCD contributions are obviously absent -- at least in the leading one-loop approximation -- from the evolution of $Y_\nu$, and the ${\rm Tr}(Y_\nu^\dag Y_\nu)$ term, which affects $\beta_\lambda$ e $\beta_{y_t}$, is always increasing. This feature, shown in the right panel of fig.~\ref{fig:RGE}, has a negative impact both on the vacuum stability and on the perturbativity of $Y_\nu$ which can be violated, if $|{Y_{\nu}}_{ij}| > \sqrt{4 \pi}$, during the RG evolution for sufficiently big values of the Yukawa coupling at the EW scale. In the same figure the decoupling of the heavy right-handed neutrinos below their mass thresholds is also manifest. Indeed, for $\mu \ll M_{Ri}$, the $N_{R}^i$ neutrino is integrated out and does not contribute to the RG running: the corresponding row in the Yukawa matrix $Y_{\nu}$ is frozen, and enters in the $\beta$ functions only above the threshold scale $M_{Ri}$ as shown in eq.~(\ref{eq:thresholds}) where we mark with a generic $\times$ non-zero entries for the Yukawa matrix $Y_{\nu}$. 
\begin{equation}
\label{eq:thresholds}
Y_\nu =
\begin{array}{ccccccc}
{\rm SM} & & {\rm EFT_1} & & {\rm EFT_2} & & {\rm Full ~ theory}  \\ \hline
& & & & & & \\
\left( \begin{array}{ccc}  0 & 0 & 0 \\ 0 & 0 & 0 \\ 0 & 0 & 0 \\ \end{array}\right) & \longrightarrow &
\left( \begin{array}{ccc} \times & \times & \times \\ 0 & 0 & 0 \\ 0 & 0 & 0 \\ \end{array}\right) & \longrightarrow &
\left( \begin{array}{ccc}  \times & \times & \times  \\ \times & \times & \times \\ 0 & 0 & 0 \\ \end{array}\right) & \longrightarrow &
\left( \begin{array}{ccc}  \times & \times & \times  \\ \times & \times & \times \\ \times & \times & \times \\ \end{array}\right) \\
& M_{R1}& & M_{R2} & & M_{R3}  & \\ 
& \mbox{\small threshold} & & \mbox{\small threshold} & & \mbox{\small threshold}
\end{array}
\end{equation}
Finally, we show in fig.~\ref{fig:BoundMeta} the evolution of the effective quartic coupling $\lambda_{\rm eff}$ in two different scenarios. In the left panel the inverse seesaw is realized with ${\rm Tr}(Y_\nu^\dag Y_\nu) \simeq 0.36$, while in the right panel the Yukawa matrix is such that ${\rm Tr}(Y_\nu^\dag Y_\nu) \simeq 0.6$. In the latter case the effects of $Y_\nu$, which affects the RG running above the threshold scales, are quite large and $\lambda_{\rm eff}$ is driven outside the metastability region below the Planck scale.

\section{Results and discussion}\label{sec:Results}

In this section we discuss the results of our analysis. First, let us briefly remind our strategy.
The Yukawa matrices generated following the prescription outlined in section~\ref{sec:NumericalSetup} 
enter as initial conditions, together with all the other SM external parameters, in the solution of the RGEs in eqs.~(\ref{eq:RGsystem1}--\ref{eq:RGsystem4}).
From the effective potential, improved by the running couplings previously computed, we extract the Higgs effective 
quartic coupling $\lambda_{\rm eff}$ in eq.~(\ref{eq:LambdaEffSm}). Finally, we use eq.~(\ref{eq:MasterBound})
to check whether the analyzed points violate the metastability bound on the EW vacuum.

\begin{figure}[!htb!]
 \begin{center}
\fbox{\footnotesize \textbf{Inverse seesaw: Normal Ordering}}
\end{center}
\vspace{-0.4cm}
\minipage{0.5\textwidth}
  \includegraphics[width=1.\linewidth]{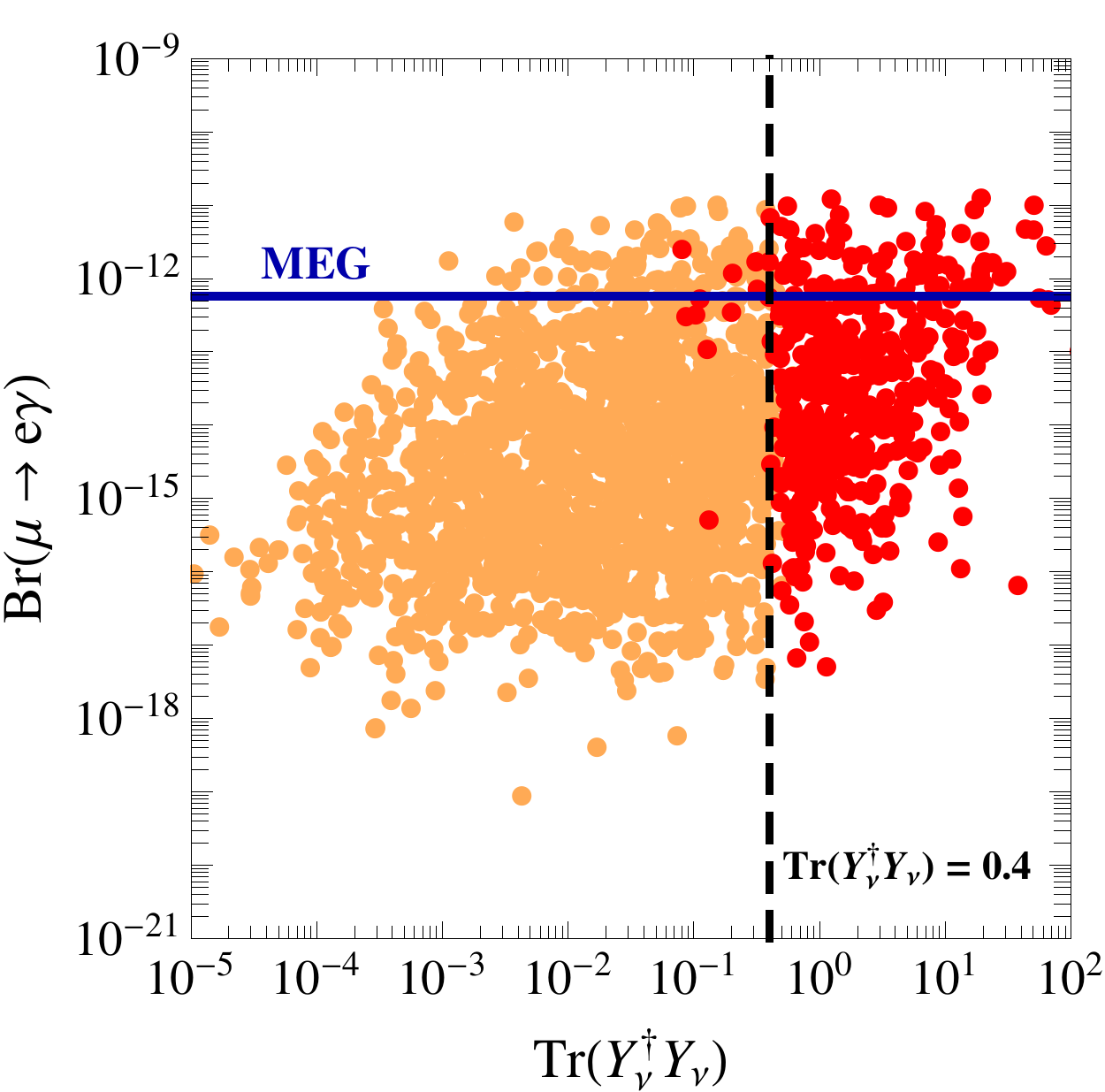}
\endminipage\hfill
\minipage{0.5\textwidth}
  \includegraphics[width=1.\linewidth]{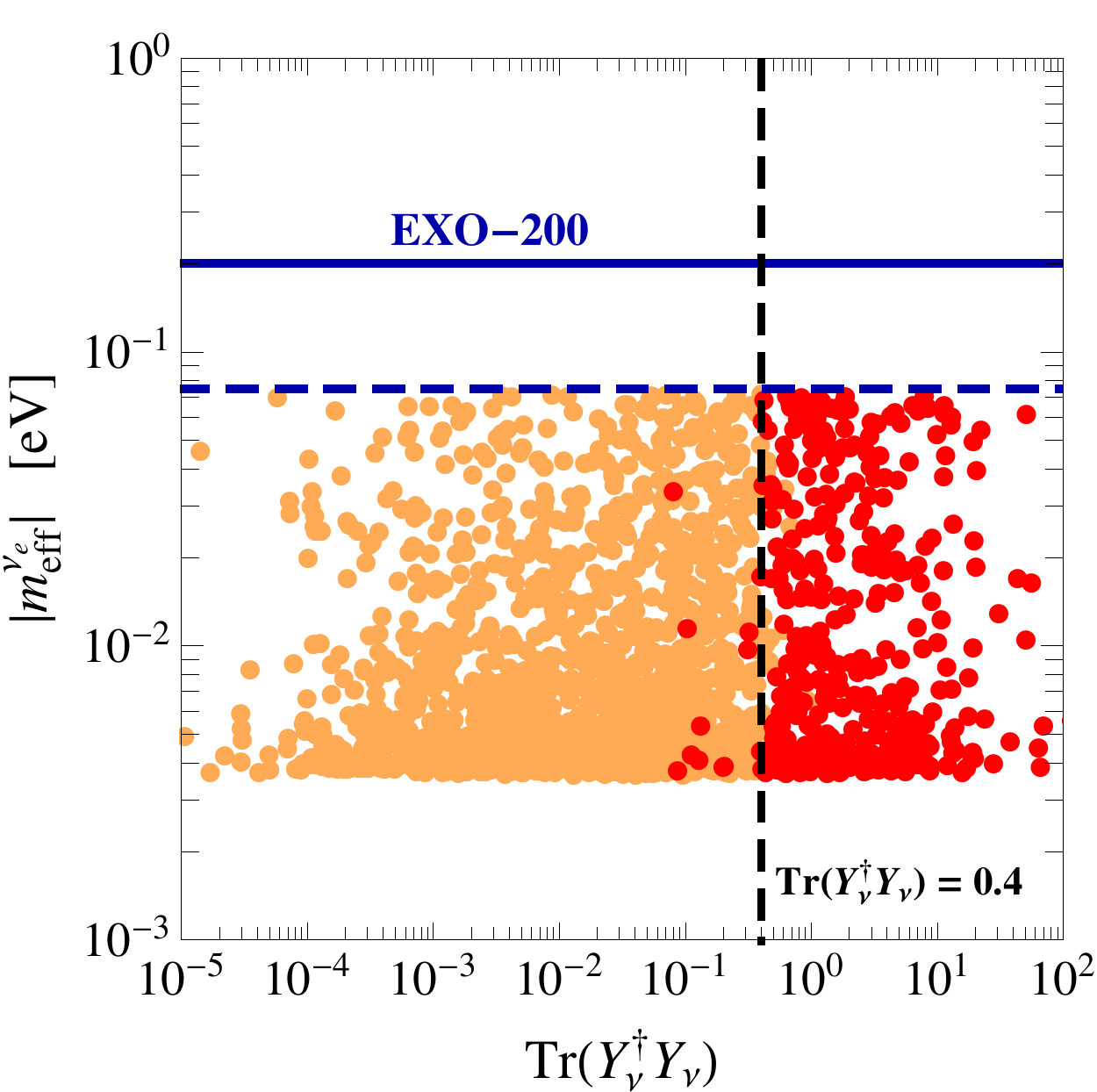}
\endminipage\\
\caption{\it 
Impact of the vacuum stability analysis on the branching ratio ${\rm Br}(\mu \to e\gamma)$ (left panel) and the effective neutrino mass (right panel) in the ISS model.
All points comply with the bounds discussed in section~\ref{sec:BoundsLowEnergy}. Red points violate the metastability bound discussed in section~\ref{sec:Stability}.
}\label{fig:FinalResultsNO}
\end{figure}

\begin{figure}[!htb!]
 \begin{center}
\fbox{\footnotesize \textbf{Inverse seesaw: Inverted Ordering}}
\end{center}
\vspace{-0.4cm}
\minipage{0.5\textwidth}
  \includegraphics[width=1.\linewidth]{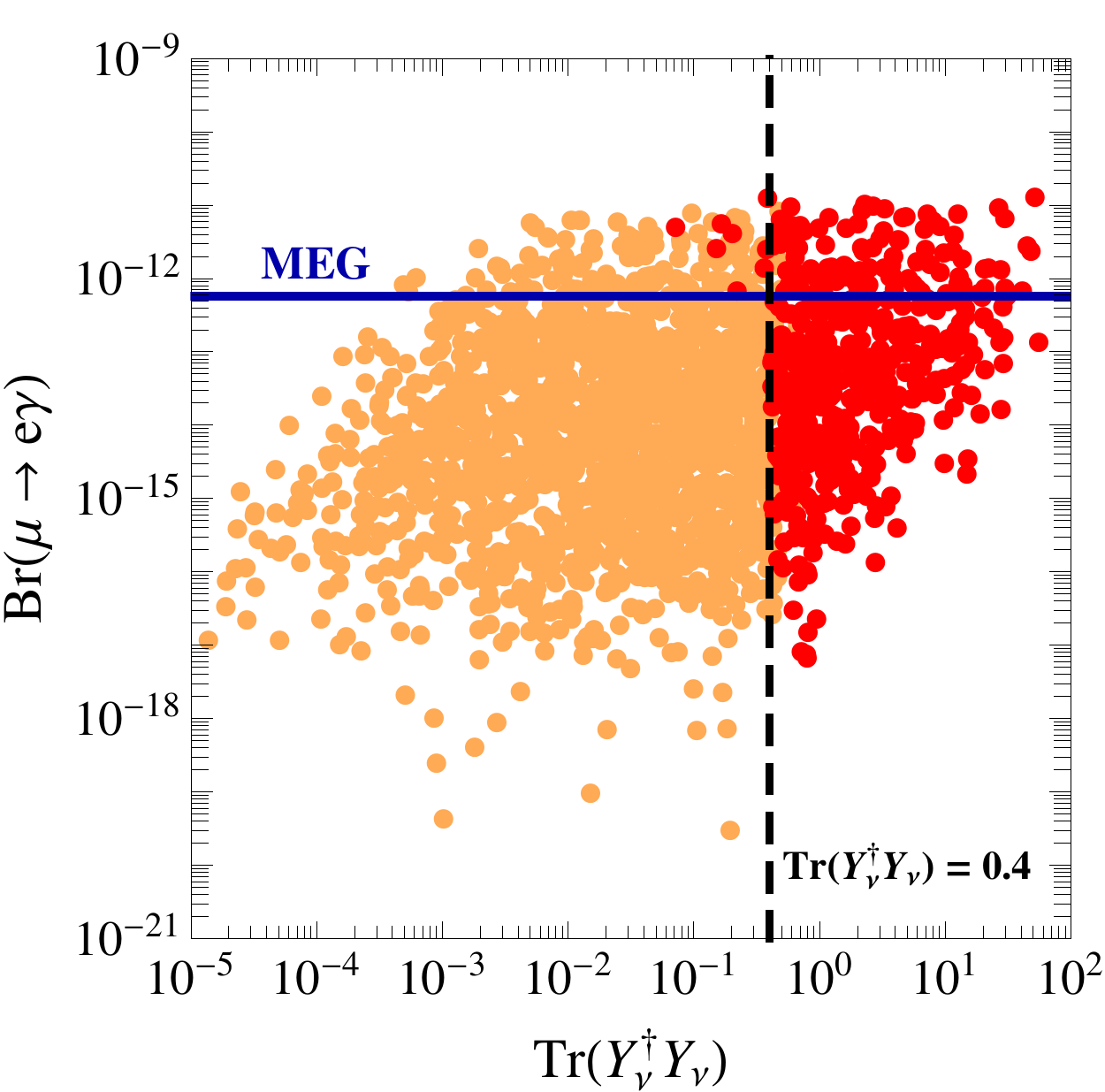}
\endminipage\hfill
\minipage{0.5\textwidth}
  \includegraphics[width=1.\linewidth]{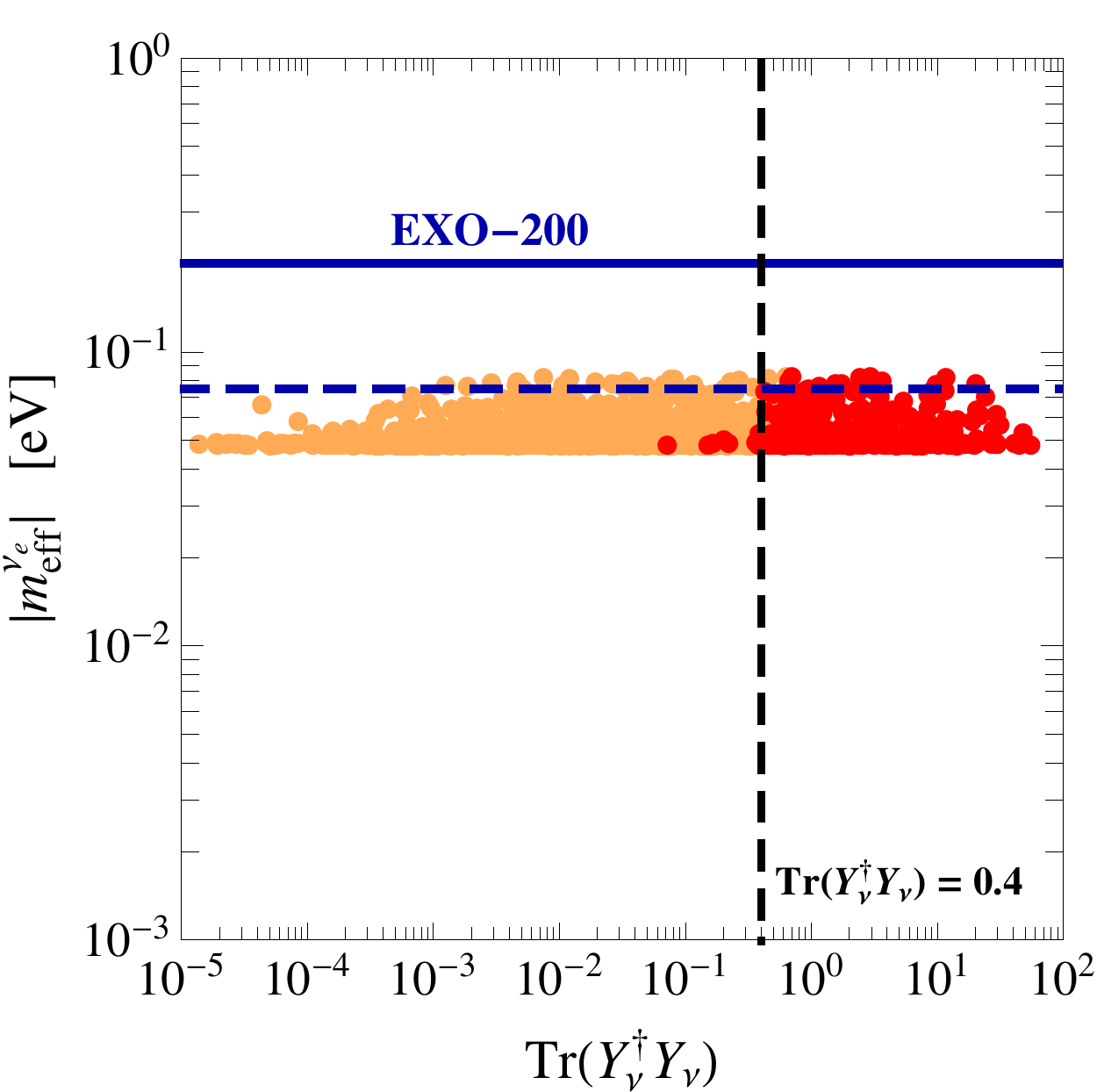}
\endminipage\\
\caption{\it 
Same as in fig.~\ref{fig:FinalResultsNO}, but considering the inverted ordering.
}\label{fig:FinalResultsIO}
\end{figure}

We show our results in fig.~\ref{fig:FinalResultsNO} for the normal ordering and in fig.~\ref{fig:FinalResultsIO}
for the inverted ordering. In order to make contact with phenomenology, in both cases we present 
the impact of the metastability bound
with respect to the observables targeted in section~\ref{sec:TargetObservables}, namely 
the branching ratio ${\rm Br}(\mu \to e\gamma)$ (left panel) and the effective neutrino mass (right panel).
The red points are excluded by the metastability bound: the EW vacuum would decay too fast in the true vacuum of the EW potential.
 
Our numerical analysis clearly indicates that points with Yukawa couplings such that  ${\rm Tr}(Y_{\nu}^{\dag}Y_{\nu}) \gtrsim 0.4$ 
are excluded. This bound does not depend on the assumed hierarchy, since light neutrino masses are irrelevant. 
Most importantly, the excluded points lie in a region of the parameter space that is
 close to the present bounds and future experimental sensitivities for both the analyzed observables.
 Moreover, as clear from the left panel of fig.~\ref{fig:Temp}, the metastability bound applies in the whole range of analyzed masses for the right-handed neutrinos. 
This is an interesting piece of information since, for instance, searches for heavy neutrinos with $m_{\nu 4} \sim \mathcal{O}(100)$ GeV and  sizable Yukawa couplings are
currently ongoing at the LHC (see section~\ref{sec:addRemarks}).

Before moving to conclusions, we refer the reader to the appendix~\ref{sec:AppA} for a generalization
of these results to the linear and double seesaw models.
In fact, one might be concerned about the model-dependence of our bound, since we worked in a well-defined model.
By extending our analysis to the two aforementioned models, 
we intend to prove that the significance of the metastability bound is a general aspect of low-scale seesaw models.

\section{Conclusions and outlook}\label{sec:Conclusions}

In this paper we analyzed the stability of the EW vacuum in the presence of low-scale seesaw models. 
For definiteness, in the main part of this manuscript we focused on the ISS model, but in appendix~\ref{sec:AppA}  we generalized our results to other popular 
low-scale scenarios.  
In the ISS model lepton number is broken at the scale $\mu_{S}$, and the scale of light neutrino masses is given by $m_{\nu} \approx \mu_S (vY_{\nu}/M_R)^2$ 
where $M_R$ is the mass of the heavy neutrinos.  
Our analysis can be divided in two parts.

First, in section~\ref{sec:Neutrino}, we carefully chose the parameter space of our numerical scan
 in order to comply with low-energy neutrino data.  We performed this selection by means of the Casas-Ibarra parametrization, and  
 we considered the intervals $10^2\,{\rm GeV}\leqslant M_{R} \leqslant 10^2\,{\rm TeV}$ and
 $10^{-1}\,{\rm keV}\leqslant \mu_S \leqslant 10^2\,{\rm keV}$
 in order to generate order one Yukawa couplings.
 As expected, the output of our numerical scan shows that the ISS model presents an extremely rich low-energy phenomenology, and we decided to focus our attention 
 on the lepton flavor radiative decay  $\mu \to e\gamma$ and on the $0\nu2\beta$.
 
Second, in section~\ref{sec:Stability}, we computed 
the contribution of the heavy neutrinos to the RG evolution of the dimensionless SM couplings and to the RG-improved effective potential. On the technical level,
we worked including the most updated results for the computation of $\beta$ functions and matching conditions in the SM, and we added the contributions of 
the right-handed neutrinos at two-loops in the  $\beta$ functions and one-loop in the matching. 

With such heavy artillery, we analyzed the Yukawa couplings generated in section~\ref{sec:Neutrino} by imposing the metastability condition on the lifetime of the EW vacuum. 
Our conclusions are summarized in fig.~\ref{fig:FinalResultsNO} and fig.~\ref{fig:FinalResultsIO}: 
Yukawa couplings such that ${\rm Tr}(Y_{\nu}^{\dag}Y_{\nu}) \gtrsim 0.4$ are excluded by the metastability condition. 
The bound applies in the whole range of analyzed right neutrino masses, 
and affects a region of the parameter space close to present or future sensitivities for both $\mu \to e\gamma$ rate 
and $0\nu2\beta$.
 
To sum up,  
we argue that the metastability bound represents an important 
consistency condition that should be included in all the phenomenological analysis of low-scale seesaw models
 featuring 
EW-scale right-handed neutrinos with $\mathcal{O}(1)$ Yukawa couplings.

\acknowledgments{

We are grateful to all the participants and organizers of the mini-workshop on ``Higgs criticality'', held at  the UZH (Zurich, May 21st, 2015), for interesting discussions.
The work of A.U. is supported by the ERC Advanced Grant n$^{\circ}$ $267985$, ``Electroweak Symmetry Breaking, Flavour and Dark Matter: One Solution for Three Mysteries" (DaMeSyFla).}

\begin{appendix}

\section{Linear and double seesaw}\label{sec:AppA}

In full generality, we introduce a Majorana mass term for the right-handed neutrinos and a Yukawa coupling for the fermionic singlets
\begin{eqnarray}\label{eq:RelevantFullLagrangian}
\mathcal{L} &=& 
i\overline{N_R}\gamma^{\mu}(\partial_{\mu}N_R) + i\overline{S}\gamma^{\mu}(\partial_{\mu}S)\nonumber\\
&-&\left[\overline{N_R}Y_{\nu}\tilde{H}^{\dag}L
+\overline{S^C}Y_S \tilde{H}^{\dag}L
 + \overline{N_R}M_R S + \frac{1}{2}\overline{N_R^{\, C}}M_N N_R +  \frac{1}{2}\overline{S^C}\mu_S S +  h.c.\right]~.
\end{eqnarray}
In the left-handed basis $N_L \equiv (\nu_L, N_R^{\,C}, S)^T$ we have, after EW symmetry breaking, the following mass matrix
\begin{equation}\label{eq:MasterMassMatrixAppendix}
\mathcal{M} =
\left(
\begin{array}{ccc}
 0  & m_D^T  & m_S^T  \\
 m_D &  M_N & M_R  \\
 m_S & M_R^T  &  \mu_S
\end{array}
\right)~,
\end{equation}
with $m_D \equiv vY_{\nu}/\sqrt{2}$ and $m_S \equiv vY_{S}/\sqrt{2}$.

\begin{itemize}

\item[$\circ$] The linear seesaw~\cite{Malinsky:2005bi} corresponds to $\mu_S =0$, $M_N = 0$
\begin{equation}\label{eq:MasterMassMatrixLinear}
\mathcal{M}_{\rm LSS} =
\left(
\begin{array}{ccc}
 0  & m_D^T  & m_S^T  \\
 m_D &  0 & M_R  \\
 m_S & M_R^T  &  0
\end{array}
\right)~.
\end{equation}
In this case the only source of lepton number violation comes from $m_S$. Following the standard seesaw approximation we find
\begin{equation}
m_{\nu} \approx m_D^T(M_R^T)^{-1}m_S + m_S^T M_R^{-1}m_D~.
\end{equation}
Generalizing the Casas-Ibarra parametrization~\cite{Casas:2001sr}, we obtain~\cite{Forero:2011pc}
\begin{equation}\label{eq:CasasIbarraLinearSeesaw}
Y_{\nu} =\frac{\sqrt{2}}{v} M_R (m_S^T)^{-1}U_{\rm PMNS}^*
\sqrt{\hat{m}_{\nu}} A \sqrt{\hat{m}_{\nu}} U_{\rm PMNS}^{\dag}~,
\end{equation}
where $A$ is a $3\times 3$  matrix satisfying the equation $A + A^T = \textbf{1}$. Consequently, $A_{ii} = 1/2$ and $A_{ij} = -A_{ji}$.

\item[$\circ$] The double seesaw corresponds to $m_S =0$
\begin{equation}\label{eq:MasterMassMatrixDouble}
\mathcal{M}_{\rm DSS} =
\left(
\begin{array}{ccc}
 0  & m_D^T  & 0  \\
 m_D &  M_N & M_R  \\
 0 & M_R^T  &  \mu_S
\end{array}
\right)~.
\end{equation}
The Majorana mass term $M_N$, in addition to $\mu_S$, violates lepton number for two units. 
We assume the hierarchy $M_N \gg M_R \gg m_D \gg \mu_S$.
Integrating out the heavy fields $N_R$, we find the effective Lagrangian
\begin{eqnarray}\label{eq:EffLDoubleSeesaw}
\mathcal{L}_{\rm DSS}^{\rm eff} &=& \frac{1}{2}\overline{L_i^{C}}\tilde{H}^*(Y_{\nu}^TM_N^{-1}Y_{\nu})_{ij}\tilde{H}^{\dag}L_j
+\frac{1}{2}\overline{S_i^C}(M_{R}^TM_N^{-1}Y_{\nu})_{ij}\tilde{H}^{\dag}L_j \\
&+& \frac{1}{2}\overline{L_i^{C}}\tilde{H}^*(Y_{\nu}^TM_N^{-1}M_{R})_{ij}S_j
+\frac{1}{2}\overline{S_i^C}(M_{R}^TM_N^{-1}M_{R})_{ij}S_j - \frac{1}{2}\overline{S^C}\mu_S S  + h.c.~.\nonumber
\end{eqnarray}
In the left-handed basis $N_L = (\nu_L, S)^T$ eq.~(\ref{eq:EffLDoubleSeesaw}) corresponds, after EW symmetry breaking, to the mass matrix
\begin{equation}
\mathcal{M}_{\rm DSS}^{\rm eff} =
\left(
\begin{array}{cc}
 -m_D^TM_N^{-1}m_D & -m_D^T M_N^{-1}M_R   \\
 -M_R^TM_N^{-1}m_D & \mu_S - M_R^T M_N^{-1}M_R 
\end{array}
\right)~.
\end{equation}
After block-diagonalization we find
\begin{equation}\label{eq:DSSdiagonalization}
V^T\mathcal{M}_{\rm DSS}^{\rm eff}V =
\left(
\begin{array}{cc}
 m_{\nu} & 0   \\
 0 & M_{\rm heavy} 
\end{array}
\right)~,
\end{equation}
where
\begin{equation}\label{eq:MLight}
m_{\nu} = -m_D^TM_N^{-1}m_D - (m_D^T M_N^{-1}M_R)(\mu_S - M_R^T M_N^{-1}M_R)^{-1}
(M_R^T M_N^{-1} m_D)~,
\end{equation}
\begin{equation}\label{eq:MHeavy}
M_{\rm heavy} = \mu_S - M_R^TM_N^{-1}M_R~.
\end{equation}
From eq.~(\ref{eq:MLight}) it follows that $m_{\nu} \to 0$ if $\mu_S \to 0$ since the type-I contribution $m_D^TM_N^{-1}m_D$ cancels out between 
the two remaining terms.
Considering a perturbative expansion in $\mu_S$, we find
\begin{equation}\label{eq:LightNeutrinosDoubleSeesaw}
m_{\nu} \approx m_D^T(M_R^T)^{-1}\mu_S M_R^{-1}m_D~.
\end{equation}
As a result, the Casas-Ibarra parametrization is analogous to eq.~(\ref{eq:CasasIbarra}).
The unitary mixing matrix $V$ in eq.~(\ref{eq:DSSdiagonalization}) has the general structure 
\begin{equation}\label{eq:MixingParameter}
V = 
\left(
\begin{array}{cc}
 \sqrt{\textbf{1} - BB^{\dag}}  &  B   \\
 -B^{\dag}  &     \sqrt{\textbf{1} - B^{\dag}B}
\end{array}
\right)~,
\end{equation}
and, at the lowest order, we find $B^* = m_D^T(M_R^{T})^{-1}$.
For the sake of simplicity, we focus on the case with $n_R = n_S = 3$, and we assume $M_R = {\rm diag}(M_{Ri})$, and $M_N = {\rm diag}(M_{Ni})$ with $i=1,2,3$. In this case the heavy block in eq.~(\ref{eq:DSSdiagonalization}) simplifies to 
$M_R^T M_N^{-1}M_R = {\rm diag}(M_{Ri}^2/M_{Ni})$. 
In eq.~(\ref{eq:EffLDoubleSeesaw}) the interactions between the lepton doublets $L_i$ and the three singlet fermions $S_i$ are mediated by an effective 
Yukawa matrix $\tilde{Y}_{\nu}\equiv M_R^T M_N^{-1} Y_{\nu}$. The previous assumption implies $(\tilde{Y}_{\nu})_{ij} = M_{Ri}(Y_{\nu})_{ij}/M_{Ni}$.

\end{itemize}

\subsection{Linear seesaw: results}

\begin{figure}[!htb!]
 \begin{center}
\fbox{\footnotesize \textbf{Linear seesaw: Normal Ordering}}
\end{center}
\vspace{-0.4cm}
\minipage{0.5\textwidth}
  \includegraphics[width=1.\linewidth]{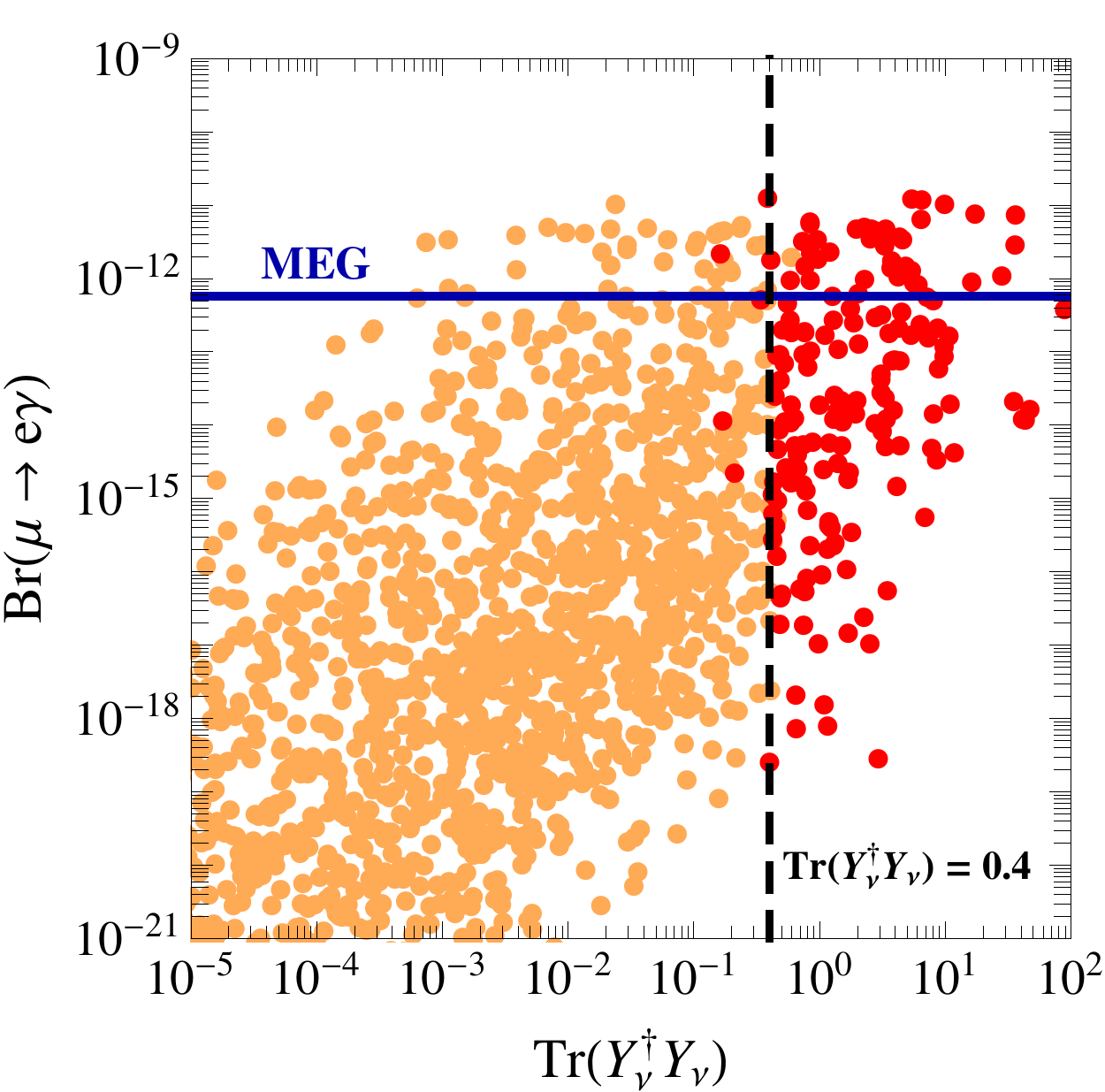}
\endminipage\hfill
\minipage{0.5\textwidth}
  \includegraphics[width=1.\linewidth]{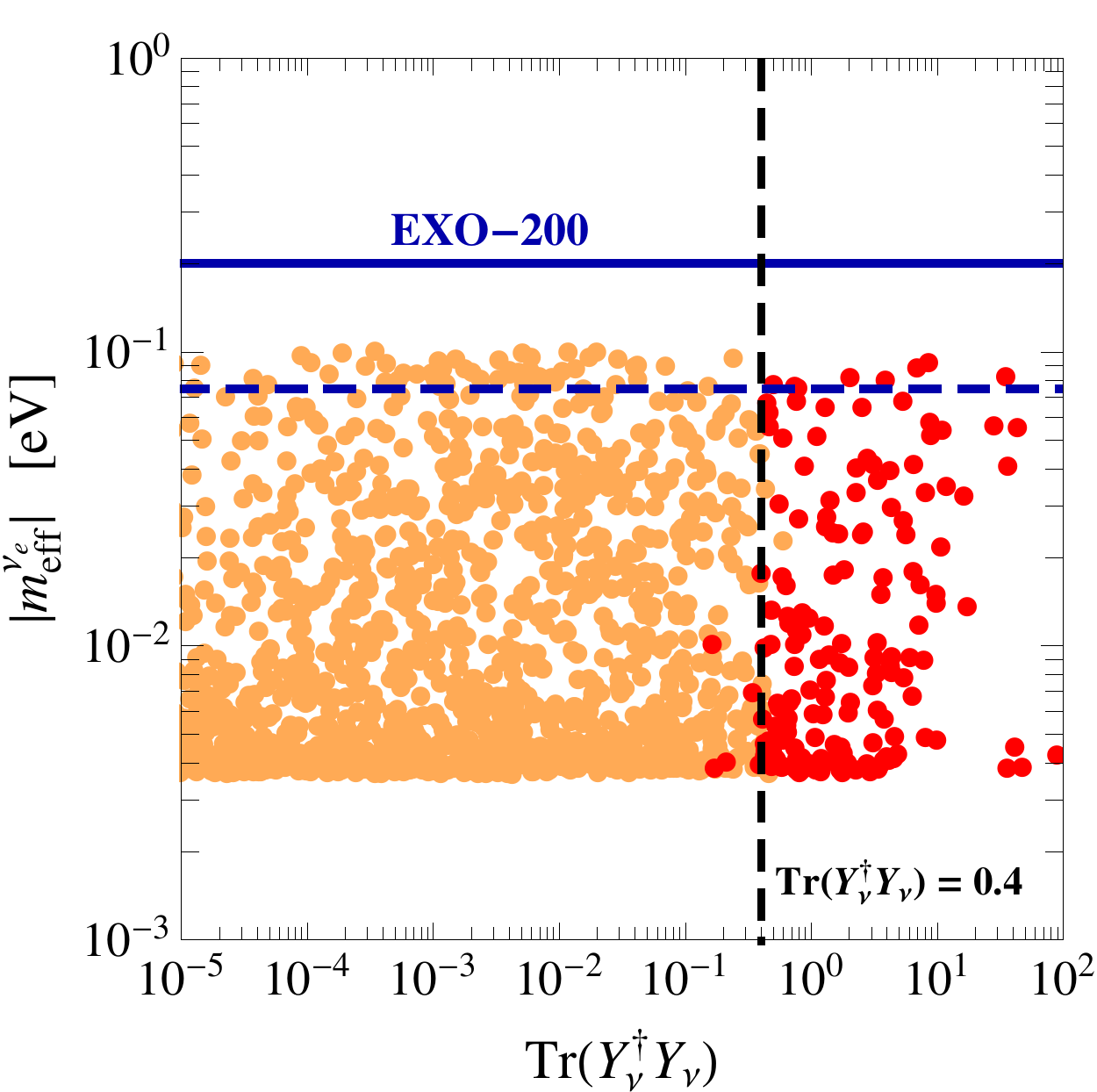}
\endminipage\\
\caption{\it 
The same as in fig.~\ref{fig:FinalResultsNO} but considering the linear seesaw.
}\label{fig:LinearSeesawFinalResultsNO}
\end{figure}

For definiteness, we focus on the case with $n_R = n_S = 3$. 
Considering the low-energy neutrino data, we follow the same strategy outlined in section~\ref{sec:Neutrino} (see in particular section~\ref{sec:NumericalSetup}).
We make use of the Casas-Ibarra parametrization in eq.~(\ref{eq:CasasIbarraLinearSeesaw}) to numerically reconstruct the 
Yukawa matrix $Y_{\nu}$, and we randomly scan over the intervals
$10^2\,{\rm GeV}\leqslant M_{R i} \leqslant 10^2\,{\rm TeV}$,
 $10^{-2}\,{\rm keV}\leqslant (m_S)_{ij} \leqslant 10^2\,{\rm keV}$, and $10^{-1} \leqslant A_{ij} \leqslant 10^2$. As done for the inverse seesaw, we discard points unable to comply with the bounds discussed in 
 section~\ref{sec:BoundsLowEnergy}.
 As far as the stability of the EW vacuum is concerned, in the $m_S \to 0$ limit the mass matrix $\mathcal{M}_{\rm LSS}$ in eq.~(\ref{eq:MasterMassMatrixLinear})
reduces to the same structure already studied in the inverse seesaw case (see eq.~(\ref{eq:MasterMassMatrix})).
Consequently, in the definition of the Higgs effective quartic coupling we employ the same RG equations, matching conditions and effective potential 
used in the inverse seesaw analysis. 
We show the final result of our analysis in fig.~\ref{fig:LinearSeesawFinalResultsNO} considering a normal ordering of light neutrino masses.

As expected, we find the same quantitative conclusion if compared with the inverse seesaw case. 
Yukawa couplings such that ${\rm Tr}(Y_{\nu}^{\dag}Y_{\nu}) \gtrsim 0.4$ 
are excluded by the metastability bound.
Remarkably, 
considering both the lepton flavor violating process $\mu \to e\gamma$ and the $0\nu 2\beta$,
these points lie in a region of the parameter space close to present or future sensitivities.
Therefore, we conclude that the metastability bound represents an important 
consistency condition that should be included in all the phenomenological analysis of the linear seesaw model featuring 
EW-scale right-handed neutrinos with $\mathcal{O}(1)$ Yukawa couplings.

\subsection{Double seesaw: results}

We follow the same approach already exploited for the inverse and linear seesaw models.
However, in the double seesaw case there are few remarkable differences.
We use the Casas-Ibarra parametrization in eq.~(\ref{eq:CasasIbarra}) to sample the Yukawa matrix $Y_{\nu}$,
and we randomly scan over the intervals 
$10\,{\rm TeV}\leqslant M_{R i} \leqslant 10^3\,{\rm TeV}$, $10^9\,{\rm GeV}\leqslant M_{N i} \leqslant 10^{11}\,{\rm GeV}$,  and  
$10\,{\rm keV}\leqslant (\mu_S)_{ij} \leqslant 10^4\,{\rm keV}$.
Few comments are in order. First, notice that this choice of parameters -- optimized in order to obtain $\mathcal{O}(1)$ Yukawa couplings -- respects the hierarchy $M_N \gg M_R \gg m_D \gg \mu_S$ assumed above (see  discussion below eq.~(\ref{eq:MasterMassMatrixDouble})).
Second, we expect the following order of magnitude estimates: for the mixing parameter in eq.~(\ref{eq:MixingParameter}), $B\sim \mathcal{O}(10^{-3})$; 
for the mass of the heavy neutrinos in eq.~(\ref{eq:MHeavy}), $M_{\rm heavy}\sim \mathcal{O}(1)$ GeV; for the effective Yukawa coupling in eq.~(\ref{eq:EffLDoubleSeesaw}), $\tilde{Y}_{\nu} \sim 10^{-5}\,Y_{\nu}$.
Armed with these numbers, we can outline as follows.
At large renormalization scale values, $\mu \gg M_{Ni}$, the model is described by the full Lagrangian in eq.~(\ref{eq:RelevantFullLagrangian}) (with $Y_S = 0$).
In terms of the RG running, the only relevant parameter is the Yukawa matrix $Y_{\nu}$ describing the interactions between the Higgs doublet, the lepton doublets and the three right-handed heavy neutrinos. At this stage, the situation is formally equivalent to the familiar type-I seesaw.\footnote{In concrete, 
the contribution of each heavy right-handed neutrino to the effective quartic coupling is 
given by eq.~(\ref{eq:NeutrinosEffPot}) (divided by two, since now there is no double degeneracy) while for the $\beta$ functions we exploit the
same two-loop expression already discussed in section~\ref{sec:Beta} (see eq.~(\ref{eq:betaynu}) for the one-loop approximation).} There is, however, one remarkable difference.
In the type-I seesaw right-handed neutrino masses $M_{N i}\sim \mathcal{O}(10^{9})$ GeV
require, in order to reproduce low-energy neutrino phenomenology, small Yukawa couplings (typically $Y_{\nu}\sim 10^{-5}$).
As a consequence, the impact of the interactions $\overline{N_R}Y_{\nu}\tilde{H}^{\dag}L +h.c.$ on the running of the Higgs quartic couplings 
is negligible. In the double seesaw case, on the contrary, we are allowed to consider $\mathcal{O}(1)$ Yukawa couplings 
since the mass of light neutrinos is set by $\mu_S$ (see eq.~(\ref{eq:LightNeutrinosDoubleSeesaw})). 
Below the thresholds $M_{N i}$ the heavy right-handed neutrinos are integrated out, and eventually the model is described by the 
effective Lagrangian in eq.~(\ref{eq:EffLDoubleSeesaw}). Given our choice of parameters, in this region the running is approximately equivalent to the 
pure SM since for the effective Yukawa interactions $\overline{L^{C}}\tilde{H}^*\tilde{Y}_{\nu}S$ we expect $\tilde{Y}_{\nu} \sim 10^{-5}\,Y_{\nu}$.

\begin{figure}[!htb!]
 \begin{center}
\fbox{\footnotesize \textbf{Double seesaw: Normal Ordering}}
\end{center}
\vspace{-0.4cm}
\minipage{0.5\textwidth}
  \includegraphics[width=1.\linewidth]{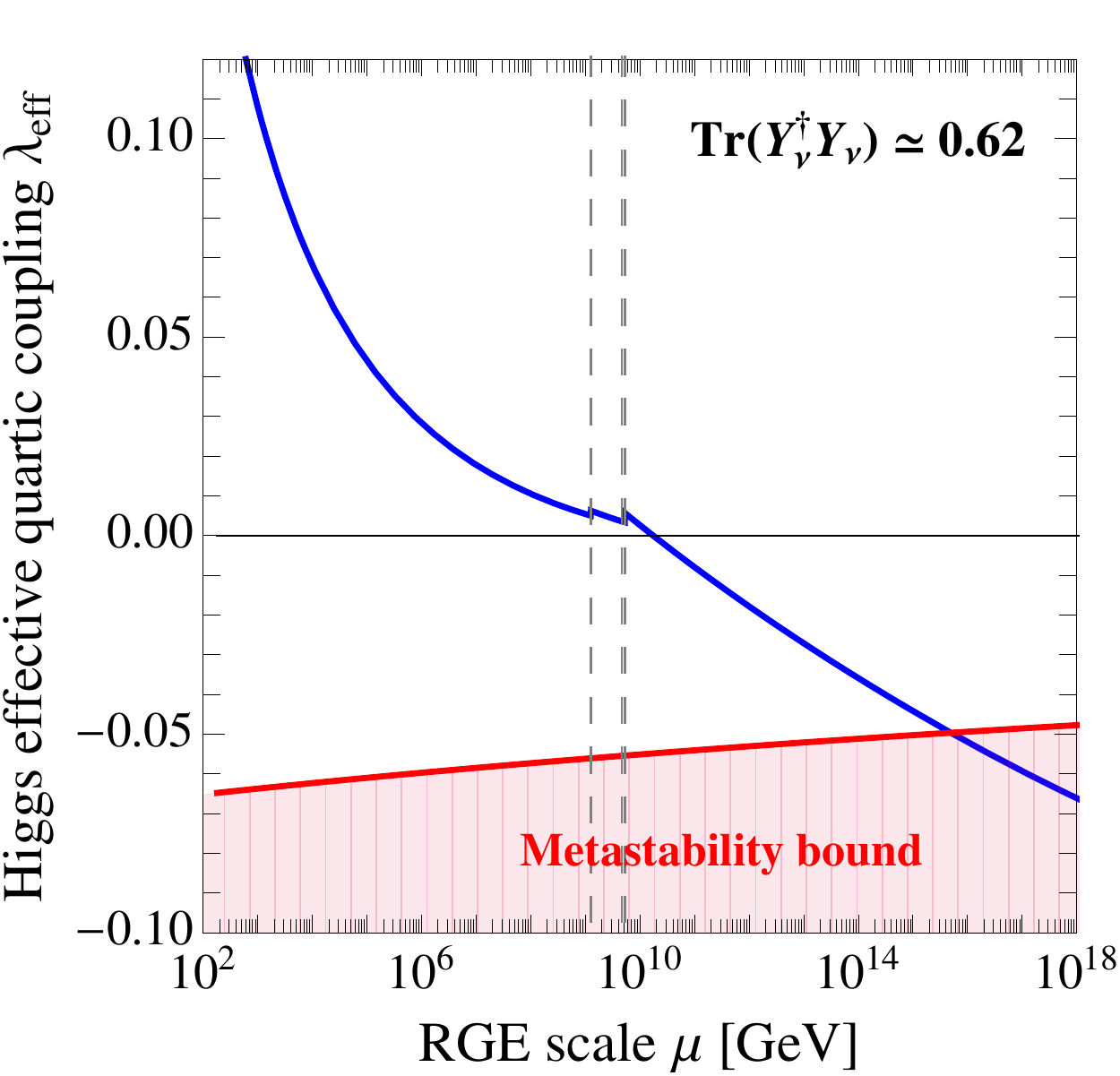}
\endminipage\hfill
\minipage{0.5\textwidth}
  \includegraphics[width=1.\linewidth]{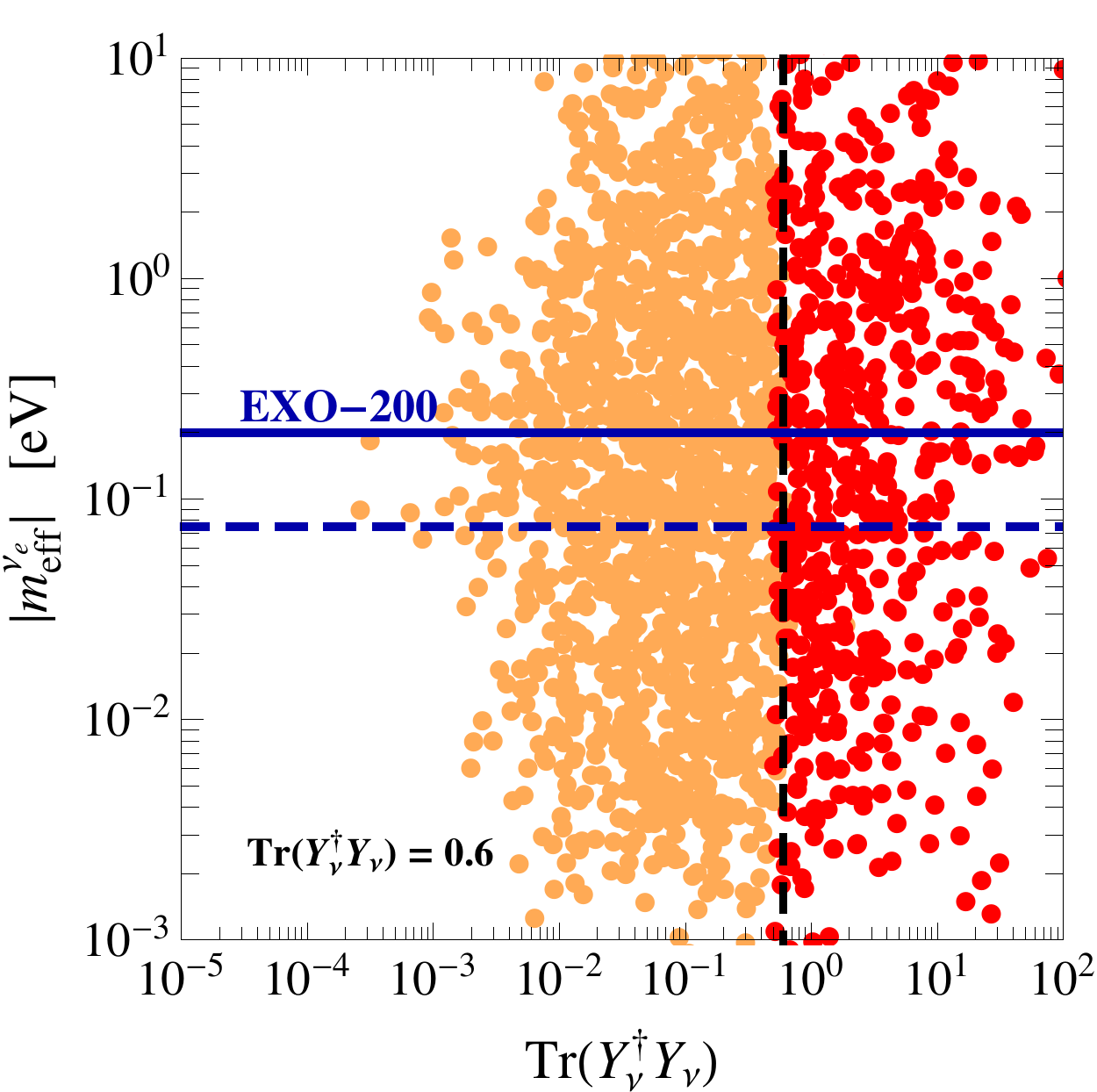}
\endminipage\\
\caption{\it 
Left panel. Running of the effective Higgs quartic coupling in the double seesaw model.
Right-handed neutrinos with mass $M_{N i}\sim \mathcal{O}(10^9)$ GeV (vertical dashed lines) sizably affect the RG evolution 
thanks to $\mathcal{O}(1)$ Yukawa couplings. In the analyzed case, Yukawa couplings such that ${\rm Tr}(Y_{\nu}^{\dag}Y_{\nu}) \simeq 0.62$ 
exceed the metastability bound below the Planck scale.
Right panel. Effective neutrino mass as a function of the 
trace
 of the Yukawa couplings, ${\rm Tr}(Y^{\dag}_{\nu}Y_{\nu})$. 
}\label{fig:DoubleSeesawFinalResultsNO}
\end{figure}

We summarize our results in fig.~\ref{fig:DoubleSeesawFinalResultsNO}.
In the left panel, we show the running of the effective Higgs quartic coupling for a specific realization of the double seesaw model 
with ${\rm Tr}(Y_{\nu}^{\dag}Y_{\nu}) \simeq 0.62$. Above the thresholds $M_{N i}$ the Yukawa couplings $Y_{\nu}$ sizably affect the running of $\lambda_{\rm eff}$
eventually violating the metastability bound before the Planck scale. 
In the right panel we show the result of our numerical scan focusing on the effective neutrino mass relevant for the $0\nu 2\beta$.
The most striking difference with respect to the inverse and linear seesaw models (see, respectively, figs.~\ref{fig:FinalResultsNO},\,\ref{fig:LinearSeesawFinalResultsNO})
is that the presence of additional neutrinos with mass  $M_{\rm heavy}\sim \mathcal{O}(1)$ GeV gives a sizable contribution to $m_{{\rm eff}}^{\nu_e}$.
As a result, numerous points in our numerical analysis are close to (or even exceed) the present experimental bound. 
We find that Yukawa couplings such that ${\rm Tr}(Y_{\nu}^{\dag}Y_{\nu}) \gtrsim 0.6$ 
are excluded by the metastability bound.

Before concluding, let us stress that the aim of this discussion is not to be exhaustive since a careful phenomenological 
analysis is clearly far beyond the scope of this appendix. 
The most important message of our discussion is that even in the presence of very heavy right-handed neutrinos (with mass
of the same size as the one expected in standard type-I scenarii) large modifications of the effective Higgs quartic couplings are a concrete and realistic possibility.
As a final remark, notice that double seesaw models represent an appealing 
setup to realize leptogenesis~\cite{Kang:2006sn,Gu:2010xc}. With this respect, a more detailed study of the parameter space will be addressed in a forthcoming publication. 

\end{appendix}


\bibliography{ref}
\bibliographystyle{jhep}

\end{document}